\documentclass[]{aa}
\usepackage[varg]{txfonts}
\usepackage{amsmath,amssymb,amsfonts,cases}
\usepackage{graphicx,color}
\usepackage{makecell}
\usepackage{natbib}

\include{aa-bib}

\usepackage{enumitem}

\newcommand{\bvec}{\mathbf}
\newcommand{\bm}{\mathbf}

\begin{document}

\title{On Flux Rope Stability and Atmospheric Stratification in Models of Coronal Mass Ejections Triggered by Flux Emergence}
\author{E.~Lee\thanks{formerly of US Naval Research Laboratory}  \and V.S.~Lukin \and M.G.~Linton}
\institute{US Naval Research Laboratory, Washington, DC 20375}
\abstract
{Flux emergence is widely recognized to play an important role in the initiation of coronal mass ejections.  The Chen \& Shibata (2000) model, which addresses the connection between emerging flux and flux rope eruptions, can be implemented numerically to study how emerging flux through the photosphere can impact the eruption of a pre-existing coronal flux rope.}
{The model's sensitivity to the initial conditions and reconnection micro-physics is investigated with a parameter study.  In particular, we aim to understand the stability of the coronal flux rope in the context of X-point collapse, as well as the effects of boundary driving in both unstratified and stratified atmospheres.}
{A modified version of the Chen \& Shibata model is implemented in a code with high numerical accuracy with different combinations of initial parameters governing the magnetic equilibrium and gravitational stratification of the atmosphere.  In the absence of driving, we assess the behavior of waves in the vicinity of the X-point.  With boundary driving applied, we study the effects of reconnection micro-physics and atmospheric stratification on the eruption.}
{We find that the Chen \& Shibata equilibrium can be unstable to an X-point collapse even in the absence of driving due to wave accumulation at the X-point.  However, the equilibrium can be stabilized by reducing the compressibility of the plasma, which allows small-amplitude waves to pass through the X-point without accumulation.  Simulations with the photospheric boundary driving evaluate the impact of reconnection micro-physics and atmospheric stratification on the resulting dynamics: we show the evolution of the system to be determined primarily by the structure of the global magnetic fields with little sensitivity to the micro-physics of magnetic reconnection; and in a stratified atmosphere, we identify a novel mechanism for producing quasi-periodic behavior at the reconnection site behind a rising flux rope as a possible explanation of similar phenomena observed in solar and stellar flares.}
{}

\titlerunning{Flux Rope Stability \& Atmospheric Stratification in Models of Coronal Mass Ejections}
\maketitle
\section{Introduction}

Coronal mass ejections (CMEs) are a common occurrence in the Sun's atmosphere that are known to release giga-tons of plasma into interplanetary space.  Some of the ejected plasma can reach the space environment of the Earth and have a strong and complex influence on space activity by inducing geospace disruptions that can severely impact spacecraft, power grids, and communication \citep{Baker13}. While CMEs are quite commonly observed \citep{Evans13}, especially during the peak of the solar cycle, they are still poorly understood.  Some of the biggest CME mysteries pertain to their origin, propagation, and relation to flares.

The initiation of CMEs has been widely studied and yet remains largely unexplained \citep[see reviews by][]{Forbes06,Chen11}.  However, many observational studies of associated features have led to clues about how they occur and what factors contribute to their destabilization \citep[see review by][]{Gopal06}.  Prior to an eruption, large-scale shear motions are often observed in photospheric images, especially about the magnetic neutral line \citep{Krall82} and in the form of sunspot rotations \citep{Tian06}.  In addition, patches of magnetic flux are found to emerge, expand, move, fragment, coalesce, and cancel over a wide range of length and time scales \citep{Sheeley69, Zwaan85, Centeno07, Parnell09}.  It is believed that shear motions, sunspot rotation, and the emergence of new flux are all related to the injection of magnetic helicity into coronal magnetic structures that could be directly involved in the eruption \citep{Chae01, Kusano02, Demoulin02, Pariat06, Magara08}.

In addition to the growing body of observational studies that have improved our understanding of CMEs, many new insights have also emerged from theoretical and numerical efforts.  CMEs have been modeled in two and three dimensions using both simple analytical methods and sophisticated magnetohydrodynamic simulations \citep[see][and references therein]{Jacobs11}.  These models differ widely in physical and numerical details, each making its own choice of how to address the trade-off between complexity and computational feasibility.

Early theoretical models explained CMEs as a loss of equilibrium, due to magnetic buoyant instabilities \citep[e.g.,][]{vanTend78, Low81, Demoulin88}, as well as MHD flows \citep{Low84} and reconnection \citep{Forbes91}.  \citet{Forbes95} proposed a CME model based on the movement of magnetic footpoints (sources) below a flux rope and the subsequent development of a singular current sheet, through which a large magnetic energy release should take place as the flux rope moves continually outwards.  \citet{LinForbes00} refined their model and computed exact solutions for the energy release, flux rope height, current sheet length, and reconnection rate.  The Lin \& Forbes (hereafter, ``LF'') model, while simplistic, provides an important step forward in CME modeling because it offers exact solutions to the time-dependent nonlinear problem of a flux rope eruption and includes more than a heuristic treatment of magnetic reconnection.  Furthermore, it predicts many features (e.g., morphology, current sheet, post-flare loops, flows, energetics) confirmed by observations \citep{Ciaravella02, Ko03, Lin05}.  

A similar two-dimensional flux rope model was proposed by \citet{ChenShibata00}.  Like LF, the Chen \& Shibata (``CS'') model consists of a two-dimensional configuration in which a flux rope sits above the photosphere, surrounded by a line-tied coronal arcade.  In both models, the magnetic equilibrium is destabilized by photospheric driving, causing a current sheet to form in the flux rope's wake as it moves outwards.  However, whereas the LF model calls for a somewhat manufactured mechanism for destabilization via large-scale convergence of the sources, the CS model improves upon the LF model by incorporating flux emergence as the driver. 

While it does not lend itself to a purely analytical treatment, the CS model is suitable for numerical simulation.  The authors report four very different outcomes based on the position and direction of the driving, showing that the location of the emergence {\em per se} is not a critical factor for destabilizing the coronal flux rope but rather that the relative orientation of the emerging flux determines whether the flux rope moves outwards/upwards (CME-like) or inwards/downwards (failed eruption).

Several subsequent studies have built upon the CS model.  For example, \citet{Chen04}, \citet{Shiota03a}, and \citet{Shiota04} produced synthetic emission images from CS simulations to compare morphological features, reconnection in-flows, and coronal dimmings found in actual CME observations.  Moreover, \citet{Shiota03a} and \citet{Shiota05} were able to identify the formation, structure, and location of slow and fast shocks in the CMEs produced in these simulations. Gravitational density stratification in an isothermal atmosphere was considered by \citet{Chen04, Shiota04} and also in a later study by \citet{Dubey06} in spherical coordinates with axisymmetry.

In this study, we re-examine the CS model using a more sophisticated numerical tool, a more realistic atmosphere, and higher spatial resolution than previous studies.  Simulations are performed using a high-order spectral element method with numerically accurate, self-consistent treatments of diffusive transport (i.e., resistivity, viscosity, and thermal conduction).  In addition, we reformulate the initial conditions to have magnetic fields that are everywhere continuous and differentiable, and to include a solar-like temperature profile with a sharp transition region and density stratification.

Through an exploration of physical parameters, we find that the CS magnetic equilibrium can be unstable even without a flux emergence driver.  Linear theory has shown that sufficient perturbation of the field lines near an X-point by waves or motion can disrupt the balance between magnetic pressure and magnetic tension, causing the X-point to collapse and form a reconnecting current sheet \citep[][chapter 2]{PriestForbes}.  Our simulations demonstrate that under a wide range of conditions the CS equilibrium is susceptible to such a collapse via nonlinear accumulation of fast magnetosonic waves at the X-point \citep{McLaughlin09}.  However, we also show that in a sufficiently incompressible plasma due, for example, to the presence of a background "guide" magnetic field co-aligned with the axis of the flux rope, the X-point collapse does not take place and the CS magnetic equilibrium can be stabilized.  For both stable and unstable configurations, we investigate the impact of the resistivity model enabling magnetic reconnection below the flux rope, as well as the plasma parameters in the low solar atmosphere, on the flux rope's response to the flux emergence driver.  We show that flux emergence can produce a rising flux rope both in a stratified and an unstratified atmosphere, though the resulting ejection speed, as well as the plasma dynamics around the X-point, can be strongly effected by the magnitude of the guide field and the atmospheric stratification.

\section{Model}
\label{sec:model}

The CS model has a two-dimensional domain with motion and magnetic field allowed perpendicular (as well as parallel) to the plane of the domain ($\bvec v, \bvec B \in \mathbb{R}^3$).  Therefore, we can write the magnetic field, normalized to some value $B_0$, in terms of a scalar potential $\psi$ representing the in-plane flux, and an out-of-plane scalar field:
\begin{equation}
  \bvec B(x,y;t) = \nabla (-\psi) \times \hat{\bvec e}_z + b_z \hat{\bvec e}_z
\end{equation}

All quantities are normalized in terms of the first three constants found in Table 1:  $L_0 = 5$~Mm, which is the unit of length; $B_0=10$~G, the unit of magnetic field strength; and $N_0 = 10^9$~cm$^{-3}$, the unit of number density.  Given the Alf\'en velocity $v_A \equiv B_0/\sqrt{\mu_0 m_p N_0}$, where $m_p$ is the proton mass, we define the unit time as $\tau \equiv L_0/v_A$, unit temperature as $T_0 \equiv B_0^2/(\mu_0 k_B N_0)$, and unit pressure as $P_0 \equiv B_0^2/\mu_0$.  The solar surface gravity $g_S = 274$~m/s$^2$ is similarly normalized as $g \equiv g_S (\tau/v_A) = 2.88 \cdot 10^{-3}$.

Due to symmetries intrinsic to the model, only half of the domain in the horizontal direction has to be resolved ($x>0$).  Thus, simulations are performed in a computational domain $(x,y)\in[0,L_x]\times[0,L_y]$, with the solar convection zone assumed to be located below the domain ($y<0$).

\begin{table}
\caption{Normalization constants}
\centering
\renewcommand{\tabcolsep}{5mm}
\begin{tabular}{c c c}
\hline
\hline
Constant & Value (MKS) & Equivalent Value\\
\hline
\hline
$L_0$ & $5 \cdot 10^6$ m & 5 Mm \\
$B_0$ & $10^{-3}$ T & 10 G \\
$N_0$ & $10^{15}$ m$^{-3}$ & $10^9$ cm$^{-3}$ \\
$v_A$ &  $6.90 \cdot 10^5$ m/s & 690 km/s\\
$\tau$ & 7.25 s & $2 \cdot 10^{-3}$ hr\\
$T_0$ &  $5.76 \cdot 10^7$ K & 4.97 keV\\
$P_0$ &  $7.96 \cdot 10^{-1}$ Pa & 7.96 dyne/cm$^2$\\
\hline
\end{tabular}
\end{table}

\subsection{Initial conditions}
\subsubsection{Magnetic configuration}

The initial magnetic configuration prescribed in the CS model consists of a coronal flux rope of radius $r_0$ surrounded by an arcade of ``loops'' that are line-tied in the photosphere.

The flux rope contains a current channel that is mirrored by an image current far below the photosphere (outside the computational domain), and four line currents produce a potential quadrupolar field just below the photosphere.  Since the bottom boundary of the numerical domain coincides with the photosphere, the only visible current initially is that within the flux rope.

In the original CS study, the coronal flux rope is given by a flux function $\psi_l$ that results in a discontinuous current density at the edge of the flux rope.  Therefore, we propose the following alternative:
\begin{subequations}
\begin{numcases}
  {\psi_l =}
    \, \dfrac{r^2}{2 r_0} - \dfrac{(r^2-r_0^2)^2}{4 r_0^3}  \ , & $r\le r_0$ \label{eq:psi_l1}\\[1mm]
    \, \dfrac{r_0}{2} - r_0 \ln r_0 + r_0 \ln r \ , & $r > r_0$ \label{eq:psi_l2}
\end{numcases}
\end{subequations}
where $r^2 = x^2 + y^2$, and the center of the flux rope lies at ($x=0$, $y=h$).

Our formulation for $\psi_l$ lends itself to a continuous current density:
\begin{equation}
j_l = - \nabla^2 \psi_l = \left\{
   \begin{array}{l @{\hspace{6mm}} c}
     \dfrac{4 r^2}{r_0^3} - \dfrac{4}{r_0}  \ , & r\le r_0 \\[4mm]
     0 \ , & r > r_0
   \end{array}
 \right.
\label{eq:current}
\end{equation}

\begin{figure}[ht]
\resizebox{\hsize}{!}
{\includegraphics[trim=10 10 10 60]{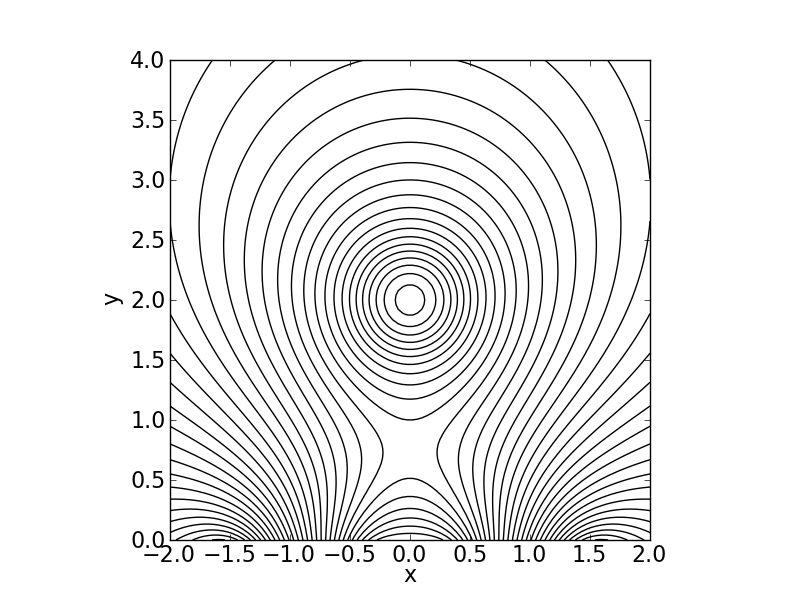}}
\caption{Contours of $\psi$ in the initial conditions.  }
\label{fig:initial_psi}
\end{figure}

The other flux components of the initial configuration, representing the image current and line currents, respectively, are kept as originally defined:
\begin{gather}
\vphantom{\int} 
  \psi_i = -\frac{r_0}{2} \ln \left[x^2 + (y+h)^2 \right]
\label{eq:psi_i}
\\[2pt]
  \psi_b = c\,\ln \frac{\left[(x+0.3)^2 + (y+0.3)^2\right]\left[(x-0.3)^2 + (y+0.3)^2\right]}
                      {\left[(x+1.5)^2 + (y+0.3)^2\right]\left[(x-1.5)^2 + (y+0.3)^2\right]}
\label{eq:psi_b}
\end{gather}
with $r_0=0.5$, $h=2$ and $c=0.25628$.
All three flux functions are summed to produce the initial magnetic equilibrium, shown in Fig.~\ref{fig:initial_psi}:
\begin{equation}
\psi = \psi_l + \psi_i + \psi_b \ .
\end{equation}

In addition to the line currents, which produce the in-plane magnetic field, we also allow for a uniform background magnetic field out of the plane $b_{z0}\,\hat{\bvec e}_z$.  This ``guide'' field contributes magnetic pressure but no current.
\begin{figure*}[ht]
\centering
\includegraphics[width=0.45\textwidth]{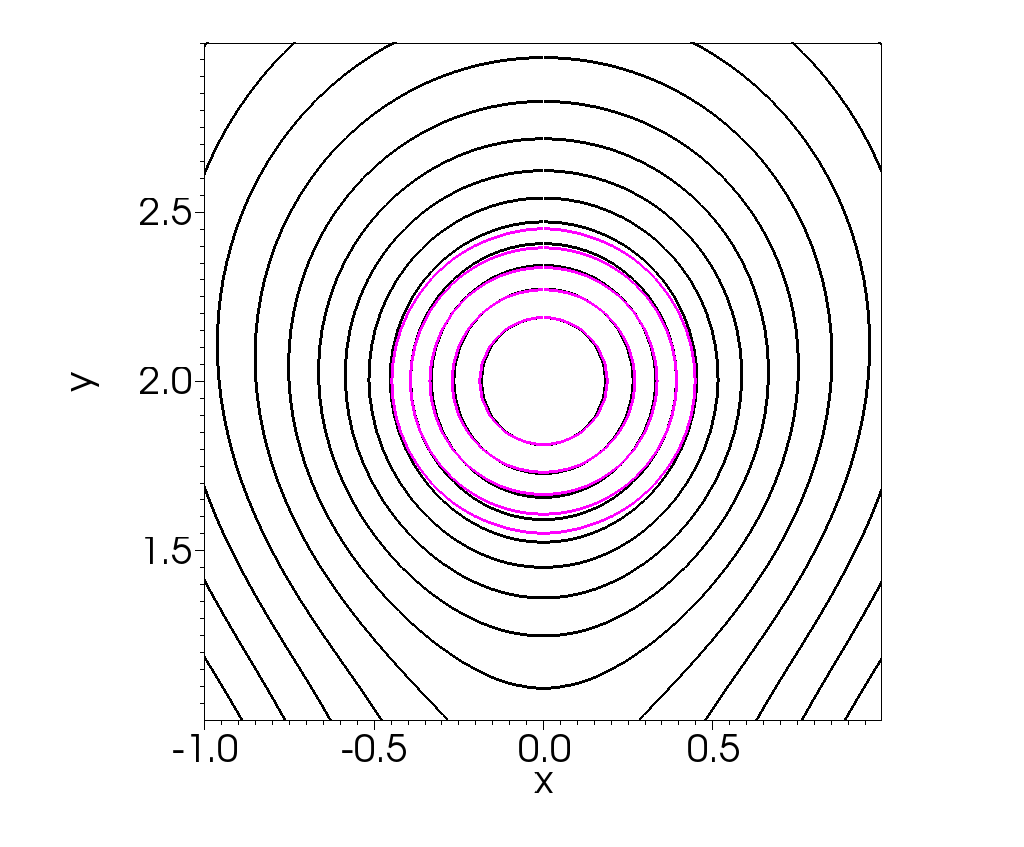}
\includegraphics[width=0.45\textwidth]{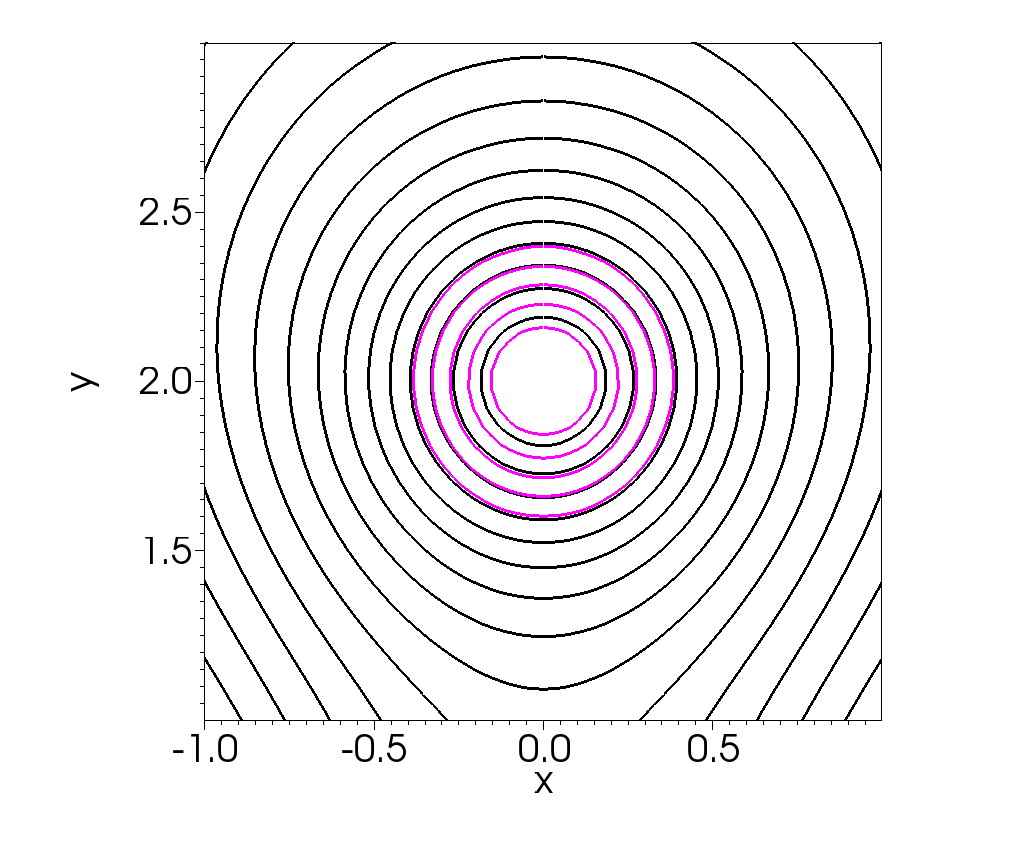}
\caption{Contours of $\psi$ (black) and $b_z$ (magenta) at $t=0$ for $b_z$ as a function of $r$ (left) and as a function of $\psi$ (right).}
\label{fig:contours}
\end{figure*}

In the original CS study, a density spike is applied to support the flux rope against radial compression:  an outward-acting pressure gradient force offsets the inward-acting Lorentz force due to the flux rope's poloidal field.  Equivalently, the flux rope can be supported against the radial Lorentz force by magnetic pressure, as in Shiota et al.~(2005):  in addition to the background guide field $b_{z0}$, we apply an additional axial field in the flux rope which is highest in the center and diminishes over a radius of $r_0$.

We note that if the axial field $b_z$ is specified as a function of the flux rope radius alone the flux rope will not be force-free, as the contours of $\psi$ are not perfectly circular due to the small but finite contributions of $\psi_b$ and, to a lesser extent, of $\psi_i$ to the total flux in the coronal flux rope.  It can be seen from the left panel of Fig.~\ref{fig:contours} that, given such a function $b_z(r)$, the contours of $\psi$ and $b_z$ would not be well-aligned.

To avoid the misalignment and minimize unbalanced Lorentz forces in the initial condition, we instead choose to specify $b_z$ as a function of $\psi$, as follows:
\begin{align}
  & b_z = \left\{ 
    \begin{array}{l @{\hspace{6mm}} c}
       \sqrt{b_{z0}^2 + \dfrac{10}{3} - 8 \left(\dfrac{\zeta}{r_0}\right)^2 + 6  \left(\dfrac{\zeta}{r_0}\right)^4 -\dfrac{4}{3} \left(\dfrac{\zeta}{r_0}\right)^6} \, & \zeta \le r_0 \\[3mm]
       b_{z0} \ , & \zeta > r_0
    \end{array}
  \right.
\label{eqn:bz_new}
\\
  & \zeta^2(\psi) = 2r_0^2 - \sqrt{3 r_0^4 - 4 r_0^3 (\psi - \psi_0)} \ .
\label{eq:rstar_approx}
\end{align}
(The derivation of the above equations can be found in Appendix A.)  Note that $\zeta=0$ when $\psi = \psi_0 - r_0/4$, and $\zeta=r_0$ when $\psi = \psi_0 + r_0/2$, where $\psi_0 \equiv \left[\psi_i + \psi_b\right] |_{(x,y)=(0,h)}$.

\subsubsection{Unstratified atmosphere}
In the case of an unstratified atmosphere, the number density field is initialized to a uniform value of $n = n_0 = 1 (N_0)$.  The pressure field of an electron-proton plasma can be determined by the following equation of state:
\begin{equation}
  p = 2 n T
\label{eq:state}
\end{equation}
We choose a uniform initial temperature, so the initial pressure $p=p_0$ is also uniform.  The free parameter $p_0$ is chosen variably in the simulations to yield temperatures close to coronal values, as well as low plasma $\beta \equiv 2p/B^2$.

An unstratified atmosphere has the advantage of isolating the flux rope dynamics from the thermodynamics.
By controlling $p_0$, one essentially explores different regimes of the plasma $\beta$.

\subsubsection{Stratified atmosphere}
\label{sec:stratified}
We also attempt to simulate a solar-like atmosphere by modeling the average vertical temperature profile as a hyperbolic tangent function, as in \cite{Leake06,Leake13}:
\begin{equation}
  T(y) = \frac{T_p}{T_0} + \left(\frac{T_c-T_p}{2 T_0}\right)\left[1+\tanh\left(\frac{y-y_\text{\tiny TR}/L_0}{\Delta y/L_0}\right)\right]
\label{eq:temperature}
\end{equation}
with photospheric temperature $T_p=5000$ K, coronal temperature $T_c=10^6$ K, transition region height $y_\text{\tiny TR}=2.5$ Mm, and transition region width $\Delta y = 0.5$ Mm.

Given this temperature profile, we seek compatible density and pressure profiles such that the plasma is in hydrostatic equilibrium:
\begin{equation}
  \frac{dp}{dy} + n g = 0
\label{eq:hydrostatic1}
\end{equation}
with the constant gravitational acceleration $g$ pointed in the $-\hat{\bvec e}_y$ direction.

We solve \eqref{eq:hydrostatic1} using \eqref{eq:state} and \eqref{eq:temperature} (see the derivation in Appendix B).  The resulting pressure profile is:
\begin{equation}
	\begin{split}
  p(y) = &p_0 \exp \left\{\frac{g\Delta y/L_0}{2 T_c/T_0} \left[ - \frac{y-y_\text{\tiny TR}/L_0}{\Delta y/L_0} 
  \right.\right.\\
  &\left.\left. 
  + \frac{T_c - T_p}{2 T_p}\ln\left(\frac{T_p}{T_0} \exp\left[- \frac{2(y-y_\text{\tiny TR}/L_0)}{\Delta y/L_0}\right] + \frac{T_c}{T_0} \right) \right] \right\}.
  	\end{split}
\end{equation}

Here the parameter $p_0$ corresponds to a constant of integration that shifts the entire pressure profile of the atmosphere.  As in the unstratified case, this affects the values of $\beta$, which should be high ($\sim$\,10) in the photosphere and low ($\sim$\,$10^{-2}$) in the corona.  Therefore, we do not vary $p_0$ for the stratified simulations.

\begin{figure}[ht]
\centering
\includegraphics[height=6cm]{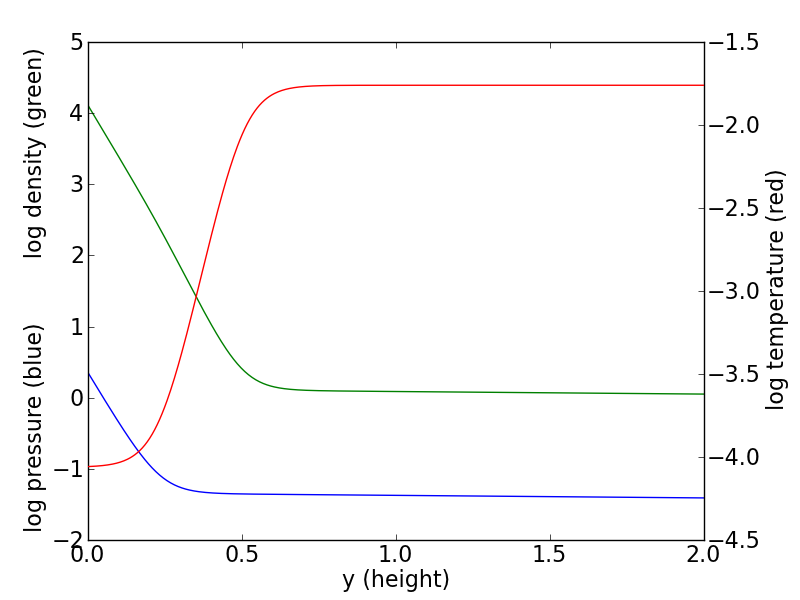}
\caption{Logarithm (base 10) of normalized number density $n$ (green), normalized pressure $p$ (blue), and temperature $T$ (red) as a function of height in the initial conditions for a stratified atmosphere.}
\label{fig:stratification}
\end{figure}

Profiles of the initial density, pressure, and temperature in a stratified atmosphere are plotted in Fig.~\ref{fig:stratification}.  It is evident that all three quantities vary smoothly and by multiple orders of magnitude.  Furthermore, this transition occurs well below the height of the flux rope core ($y=2$).

\subsection{Numerical method}

We implement this initial configuration in the high-fidelity numerical simulation
framework, HiFi, which makes use of high-order spectral elements and implicit
time-stepping \citep{LukinThesis,Lukin11}.  As a strong condition, HiFi requires all variables to be represented by continuous functions in the initial and boundary conditions.  Therefore, the magnetic field must be everywhere differentiable, implying $\psi \in \mathcal{C}^2$.  In addition, the boundary driving of flux needs to be differentiable in both space and time, in order for the electric and magnetic fields to be smooth.  These conditions are well satisfied by the initial conditions described above.

In this work, HiFi is used to integrate in time the following equations of visco-resistive MHD:
\begin{subequations}
\begin{gather}
  \frac{\partial n}{\partial t} + \nabla \cdot \left( n \bm v \right) = 0
\label{eq:density}
\\
  \frac{\partial (-\psi)}{\partial t} = - \bvec v \times \bvec B + \eta j_z
\label{eq:psi}
\\
  \frac{\partial b_z}{\partial t} + \nabla \cdot \left( b_z \bvec v - v_z \bvec B \right) = \nabla \cdot \left( \eta \nabla b_z \right)
\label{eq:bz}
\\
  \frac{\partial n \bvec v}{\partial t} + \nabla \cdot \left\{ n \bvec v \bvec v + p \bvec I - \mu n \left[\nabla \bvec v + \left(\nabla \bvec v \right)^T\right]\right\} = \bvec j \times \bvec B
\label{eq:momentum}
\\
  \frac{3}{2} \frac{\partial p}{\partial t} + \nabla \cdot \left( \frac{5}{2} p \bvec v - \kappa \nabla T \right) = \bvec v \cdot \nabla p + \eta j^2 + \mu n \left[\nabla \bvec v + \left(\nabla \bvec v \right)^T\right]:\nabla \bvec v
\label{eq:pressure}
\end{gather}
with an auxiliary equation (Amp\`ere's law):
\begin{equation}
\nabla \times \bvec B = \bvec j
\end{equation}
\label{eq:mhd}
\end{subequations}
The normalized transport coefficients found in Eqs.~\eqref{eq:mhd} -- namely $\kappa$, $\mu$, $\eta$ -- control the level of dissipation of the MHD fluid quantities through molecular diffusion: temperature, velocity, and current, respectively.  Each of these three transport parameters is chosen to be compatible with the resolution and objective of each simulation.  Further, for some of the simulations (see below), we allow the resistivity $\eta$ to be a function of local current density, $\eta = \eta_{bg} + \eta_{anom}(\bvec j)$, where
\begin{subequations}
\begin{numcases}
  {\eta_{anom}(\bvec j) =}
    \, 0 & $|\bvec j| < j_c$ \nonumber \\
    \, \bar{\eta}_{anom}\frac{\left\{1 - \cos\left[\pi(|\bvec j|/j_c - 1)\right]\right\}}{2} & $j_c \le |\bvec j| \le 2j_c$,  \nonumber \\
    \, \bar{\eta}_{anom} & $ |\bvec j| > 2j_c$ \nonumber
\end{numcases}
\end{subequations}
$\eta_{bg}$ is the uniform and time-independent background resistivity, and $\eta_{anom}$ is some ``anomalously enhanced" effective resistivity, $\bar{\eta}_{anom} \gg \eta_{bg}$, occurring due to micro-physics not captured by the MHD model whenever the current density rises above the critical current density $j_c$.


\subsubsection{Boundary conditions}
\begin{table*}
\caption{Boundary conditions}
\centering
\renewcommand{\tabcolsep}{8mm}
\renewcommand{\arraystretch}{1.5}
\renewcommand\cellgape{\Gape[5pt]}
\begin{tabular}{c c c}
\hline
\hline
Boundary & Unstratified & Stratified \\
\hline
\hline
Left (reflection) & only vertical flow & only vertical flow \\
\hline
Top/Right (coronal) & \makecell{only vertical flow,\\
                      $\nabla_{\hat{n}} \{n, b_z, j_z, p\} = 0$}
                          & \makecell{only vertical flow,\\
                            $\nabla_{\hat{n}} \{n, b_z, j_z, p\} = 0$} \\
\hline
Bottom (photosphere) & \makecell{no flow, \\
                      $\nabla_{\hat{n}} \{n, b_z, j_z, T\} = 0$}
                          & \makecell{only out-of-plane flow, \\
                            $\partial_t\left[\nabla_{\hat{n}}\{\ln(n)\}\right] = 0$, \\
                            $\nabla_{\hat{n}} \{nv_z, b_z, j_z, T\} = 0$} \\
\hline
\end{tabular}
\label{table:BC}
\end{table*}
The bottom boundary of the simulation domain, representing the photosphere, is perhaps the most important boundary condition affecting the outcome of a simulation.  Flux emergence is achieved by varying the flux function at this boundary in time, which is equivalent to applying an electric field.  This electric field determines the evolution of the magnetic field, which can be advected in or out of the domain or resistively dissipated, as described by Ohm's Law:
\begin{equation}
  \frac{\partial \psi}{\partial t} = E_z = - \hat{\bvec e}_z \cdot \bvec v \times \bvec B + \eta j_z
\label{eq:Ohm}
\end{equation}
The resistive component, the second term on the right-hand side of \eqref{eq:Ohm}, is determined by the geometry of the magnetic field at any given time.  Therefore, by varying the flux at the boundary ($\partial \psi / \partial t$) in a prescribed way, we also induce cross-field plasma motions ($\bm v \times \bm B$).

\citet{ChenShibata00} prescribe two cases of localized boundary driving, namely, over a region $|x-x_0| \le 0.3$ centered at $x_0=0$ (case A) and at $x_0=3.9$ (case B).  We apply the same method only for the case of $x_0=0$, but use a formulation that is smoother in time and in space:
\begin{equation}
  \begin{array}{c c}
\psi(x,0;t) = \psi(x,0;0) + \dfrac{\psi_e(x)}{2} \left[\dfrac{t}{t_e} - \dfrac{\sin \left(2 \pi t/t_e\right) } {2 \pi} \right] \ , & t \le t_e \\[5mm]
\psi_e(x) = \dfrac{c_e}{2}\left[ 1 + \cos \dfrac{\pi(x-x_0)}{0.3}\right] \ , & |x|\le 0.3
  \end{array}
\label{eq:driving}
\end{equation}
where $t_e$ is the duration over which the electric field drive is applied at the boundary.  For $t > t_e$, the photospheric boundary is treated as a perfect conductor. 

We do not allow any in-plane flow on the bottom photospheric boundary ($v_x = 0$, $v_y=0$) and force the normal gradients of $b_z, j_z$, and temperature to be zero:  i.e., $\nabla_{\hat{n}} \equiv \hat{n} \cdot \nabla = 0$.  The unstratified atmosphere also has $\nabla_{\hat{n}} n = 0$, while the stratified case imposes a fixed value of the density scale height $\partial_t\left\{[\nabla_{\hat{n}} n]/n\right\} = \partial_t\left\{\nabla_{\hat{n}}[\ln(n)]\right\} = 0$.

The left boundary is a symmetry boundary, with odd symmetry required for the horizontal and out-of-plane components of flow, $v_x$ and $v_z$, and even symmetry imposed on all other dependent variables.  At the outer boundaries (top and right), the gradients of density, $b_z$, $j_z$, and pressure are zero, and flow is only allowed in the vertical direction.  Table \ref{table:BC} provides a simple reference for the various boundary conditions applied in the simulations.

\subsubsection{Dissipative Boundary Layers}

\paragraph{Chromosphere.}  The flux emergence represented by \eqref{eq:driving} changes the flux function just at the boundary but has no direct effect on $\psi$ anywhere else, including just above it.  While Ohm's Law \eqref{eq:Ohm} does relate flux evolution to fluid transport, it does not guarantee that the flux function and other quantities will be well-behaved for all time, particularly in the $y$-direction.  Therefore, to allow flux to slip more easily through the region just above the photospheric driving (effectively, the ``chromosphere''), we apply a resistive boundary layer by enhancing $\eta$ locally according to the following function:
\begin{equation}
  \eta = \eta_{bg} + \eta_{anom} + \eta_{ph} \exp\left(-\frac{y^2}{y_0^2}\right)
\end{equation}
where $\eta_{ph}$ is the photospheric value of resistivity, and $y_0=0.2$ corresponds to the height of 1~Mm above the photosphere.  Conceptually this boundary layer emulates the enhanced resistivity of the chromosphere due to collisional impedance by neutrals \citep{Leake06}.

\paragraph{Outer corona.}  Separately, to mimic an open domain boundary with no wave reflections in the corona, a viscous boundary layer is prescribed close to the outer coronal (top and right) boundaries.  Starting at a distance of $d=0.5$ inward from the boundary of the computational domain, $\mu$ is increased gradually towards the domain boundary (according to a cosine profile) from a background value of $\mu_{bg}$ up to the outer boundary value of $\mu_{out}=1$.  In the same fashion, near the outer boundaries with $d=0.5$, resistivity is ramped up from $\eta_{bg}$ to a boundary value of $\eta_{out}=10^{-2}$.

\section{Results}
\label{sec:results}

In this section, we discuss simulations of the undriven equilibrium, as well as driven simulations of flux emergence in both the stratified and unstratified cases.  While the CS initial conditions describe an approximate equilibrium, we find that this equilibrium can be unstable to small perturbations.  We discuss the role of MHD waves in destabilizing the flux rope via X-point collapse and the role played by plasma compressibility in stabilizing the X-point and the flux rope.  Finally, we discuss the driven simulations of stable and unstable equilibria with different resistivity models, as well as details of the resulting eruption process.
\begin{figure*}[ht]
\centering
  \includegraphics[trim=0 0 0 0,clip,height=5.8cm]{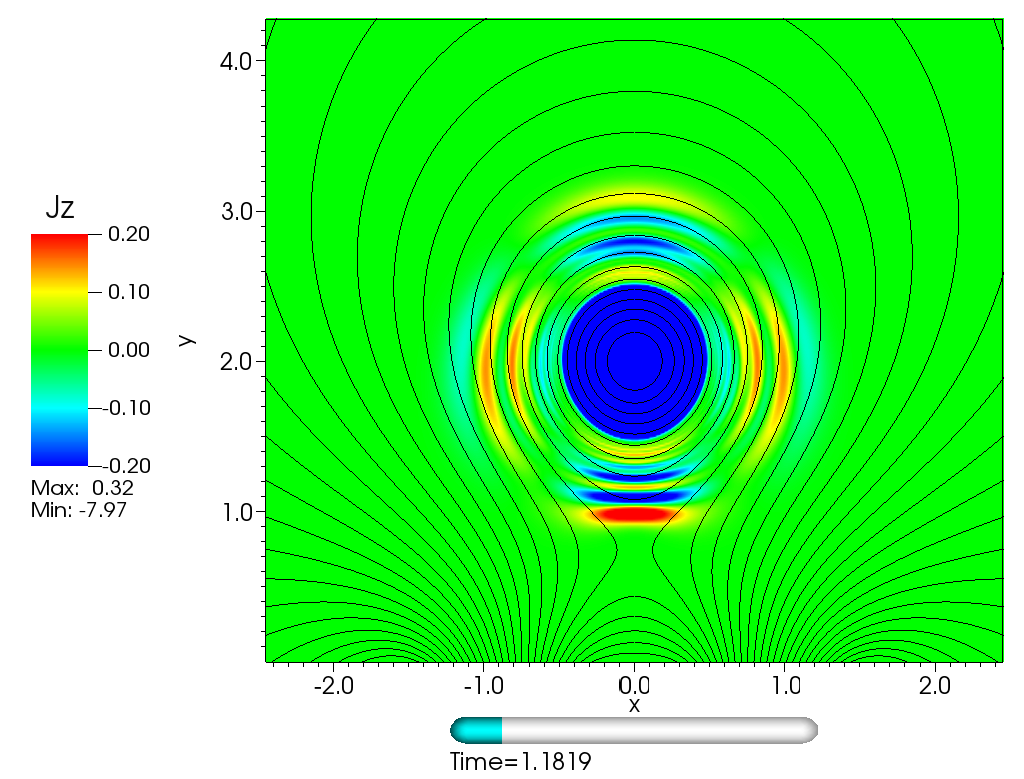} 
  \includegraphics[trim=0 0 0 0,clip,height=5.8cm]{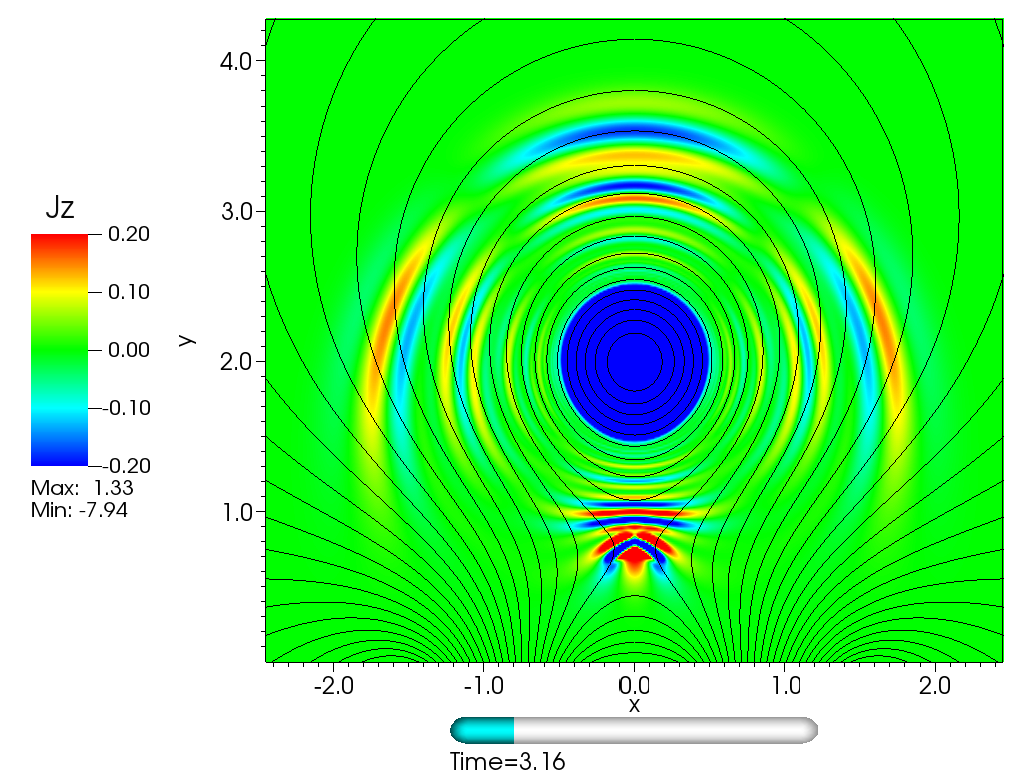}
  \includegraphics[trim=0 0 0 0,clip,height=5.8cm]{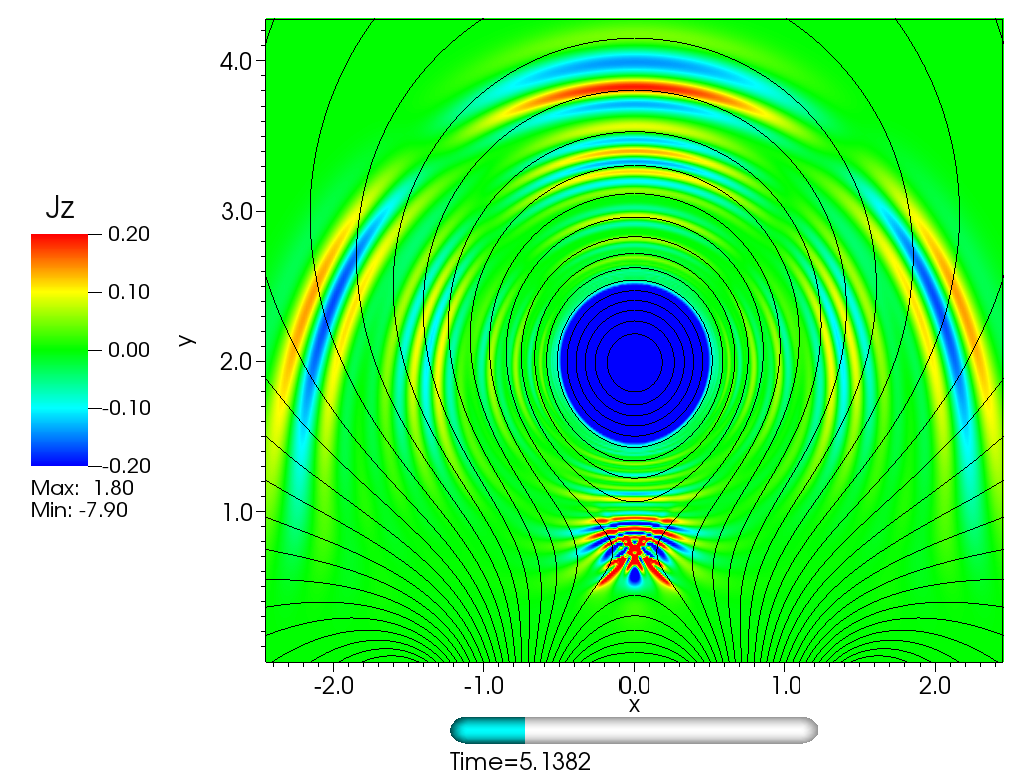}
  \includegraphics[trim=0 0 0 0,clip,height=5.8cm]{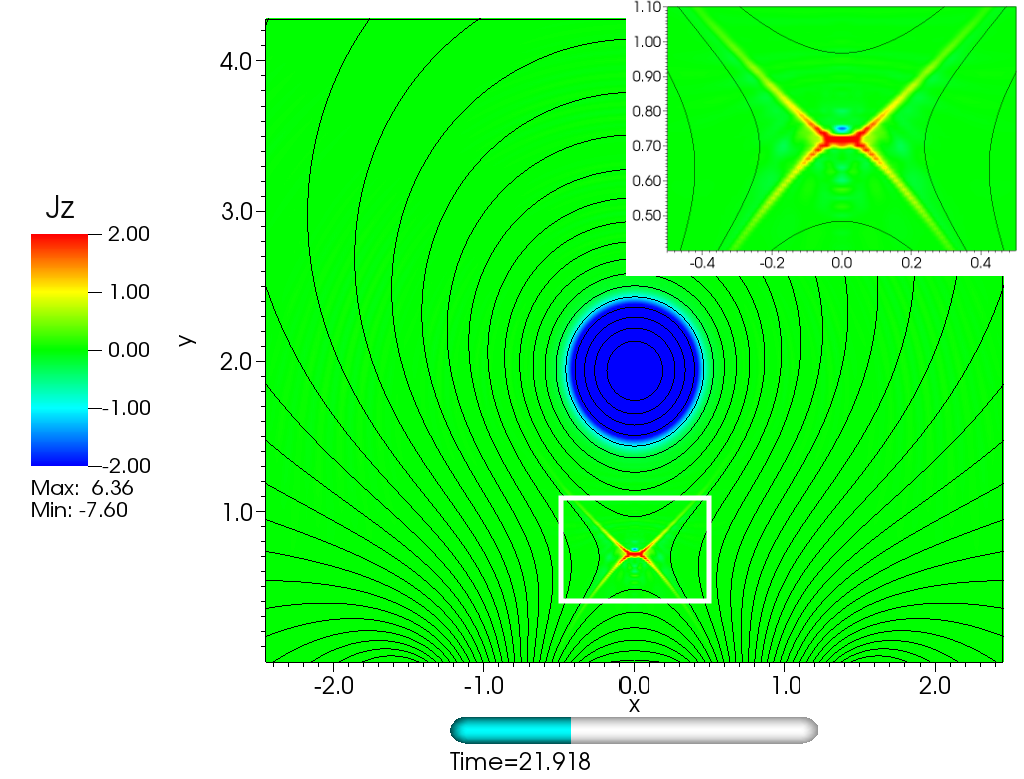}
\caption{(\textbf{No driving.}) Four snapshots of current density $j_z$ showing outward propagating MHD waves.  Notice the trapping and interference of waves at the X-point; compounding of these waves here precipitates an X-point collapse, leading to the formation of a current sheet and initiation of magnetic reconnection. 
}
\label{fig:waves}
\end{figure*}

\subsection{Flux rope stability}

In \citet{ChenShibata00}, the coefficient $c$ (see Eq.~\ref{eq:psi_b} above) was determined by trial and error to yield a magnetic equilibrium such that the center of the flux rope did not move ``for long enough time.''  (It can also be derived by requiring that the vertical component of the Lorentz force is zero at the center of the flux rope, $\hat{y}\cdot[\bvec j\times\bvec B]|_{(x,y)=(0,h)}=0$.)  Our numerical experiments confirm that this value for $c$ is indeed appropriate for equilibrium, but we find that the equilibrium itself is tentative and unstable to perturbations.

\subsubsection{X-Point collapse}

At the beginning of each simulation, the flux rope---which is approximately force-free---goes through a small adjustment to settle into an actual force-free equilibrium.  In Sec.~\ref{sec:model}, we explained how formulating $b_z$ as a function of $\psi$ forces the contours of $b_z$ and $\psi$ to be aligned.  Although this is an improvement on the original setup, the configuration is still not perfectly force-free because the contours of $j_z = -\nabla^2 \psi = -\nabla^2 \psi_l$ are still circular and therefore not aligned with the contours of $\psi$, producing a small but finite Lorentz force.  The adjustment to correct the misalignment, however small it may be, is sufficient to generate waves that may destabilize the flux rope.

Fig.~\ref{fig:waves} illustrates the oscillation of current density induced by the fast magnetosonic waves emanating from the flux rope as it adjusts to the initial conditions.  The distribution of the waves is not uniform because the initial adjustment is one where the flux rope is squeezed in one direction (horizontally) while expanding in the other direction (vertically).  Therefore, the waves propagating horizontally are out of phase with those propagating vertically.  It is interesting to note that the propagation of both types of waves is initially radial but eventually becomes oblique at the flanks of the active region due to the inhomogeneity of the magnetic pressure in the corona.  

The fast waves themselves do not directly destabilize the flux rope.  However, the entire equilibrium can be destabilized when the waves reach the X-point below the flux rope and cause it to collapse.  The role of fast waves in X-point collapse for a zero-$\beta$ plasma has been studied by \citet{McLaughlin09} and their behavior in the neighborhood of X-points has been investigated by similar earlier studies \citep{McLaughlin04, McLaughlin05, McLaughlin06}.  As described in these studies, we find that the fast waves approach the X-point but tend to get trapped there if their phase speed becomes too low at the X-point.  The trapping occurs because the waves are refracted towards the null and then wrap around it if they cannot pass through it.  As a result, the waves push current towards the null where it accumulates exponentially in the linear regime.  The buildup of waves at the null, however, quickly leads to nonlinear behavior, forming shocks and jets, which deform the X-point into a cusp-like geometry, which flattens and forms a current sheet \citep{McLaughlin09}.  This fast-wave accumulation and resulting collapse of the X-point are evident in the sequence of figures in Fig.~\ref{fig:waves}.

The consequence of X-point collapse occurring as a result of fast wave accumulation at the null is that it forms a current sheet separating anti-parallel fields.  The formation of the current sheet, in turn, kicks off magnetic reconnection that drives itself for as long as there is free magnetic energy available in the system.  When the collapse forms a horizontal current sheet, the reconnection process draws in the flux rope from above, and it pulls itself down towards the photosphere to draw in flux from below, destroying the original configuration.  Formation of a vertical current sheet similarly leads to a CME eruption.

Other factors that may contribute to a collapse of the X-point in the absence of driving include boundary flows, likely related to the reflection of waves, and the asymmetric resistivity model, which intentionally biases $\eta$ in the $y$-direction in order to allow magnetic flux to slip through the photosphere (see previous section).  However, we found these effects to be sub-dominant to the fast wave accumulation at the X-point in destabilizing the flux rope.
\begin{figure*}[ht]
\centering
\includegraphics[height=6.5cm]{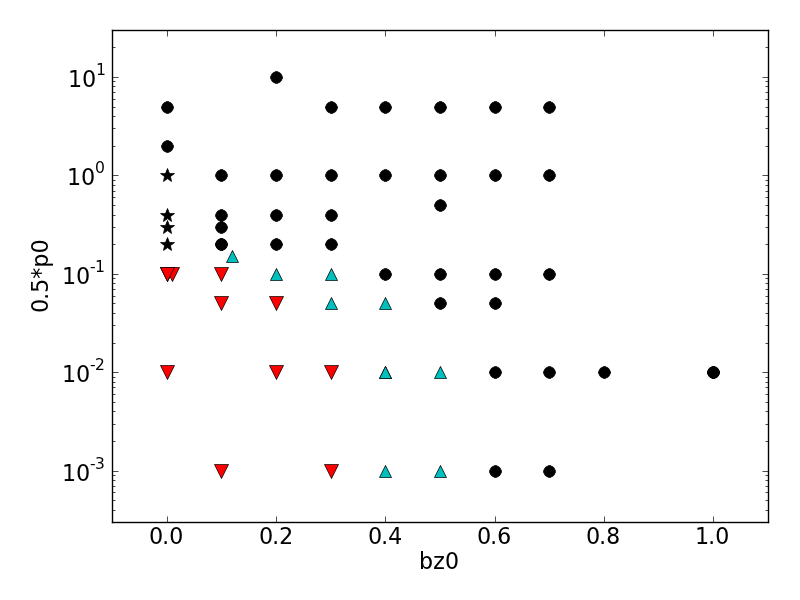}
\includegraphics[height=6.5cm]{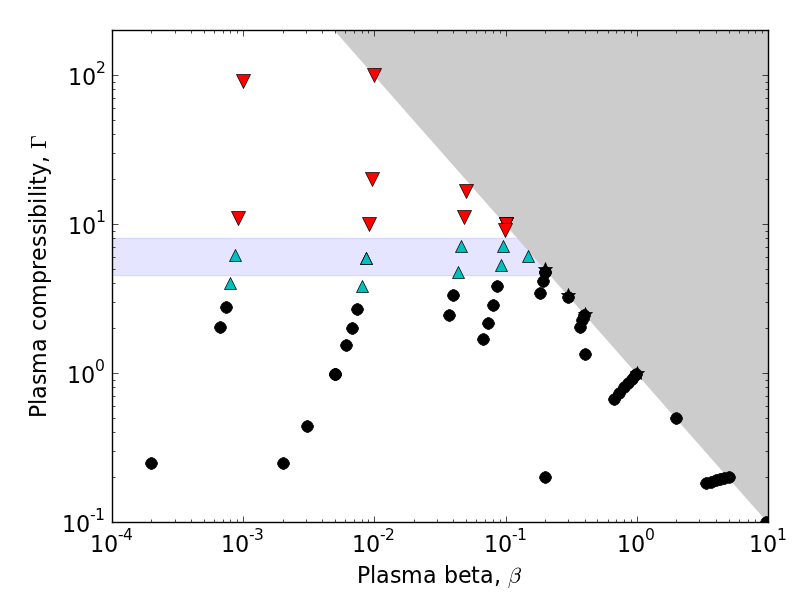}
\caption{Dependence of flux rope stability on the free parameters $p_0$ and $b_{z0}$ (left), as well as on the plasma $\beta$ and compressibility measure $\Gamma$ (right).  All simulations are performed {\em without} driving.  The different symbols signify that the flux rope was stable (black dots); was wobbly but on average did not rise or sink (black stars); moved upwards (blue triangles); or moved downwards (red triangles).}
\label{fig:sensitivity}
\end{figure*}
\subsubsection{Sensitivity to Compressibility}

We find that sufficient magnetic pressure and/or gas pressure at the X-point can suppress the X-point collapse.  In the absence of a guide field ($b_{z0}=0$), the magnetic pressure drops to zero at the X-point.  Then, with low gas pressure, the fast wave speed is reduced drastically at the X-point causing wave refraction and accumulation.  However, a parametric exploration of $p_0$ and $b_{z0}$ in an unstratified atmosphere revealed that increasing either of these parameters helps to stabilize the flux rope.  The left panel of Fig.~\ref{fig:sensitivity} is a graphical chart of the many simulations that were performed scanning the parameter space of $p_0$ and $b_{z0}$.  Black dots represent simulations in which the flux rope was stable over many hundreds of Alfv\'en times (no X-point collapse); blue triangles represent those in which the flux rope experienced a slow rise (vertical collapse); red triangles represent those in which the flux rope descended (horizontal collapse); and black stars represent those in which the flux rope moved up and down but on average maintained the same height in the atmosphere (oscillatory X-point collapse).

One could argue heuristically that increasing either $p$ or $b_z$ effectively decreases the compressibility of the plasma (or increases the stiffness of the medium), so any motions at the X-point need to do more work against the gas pressure or magnetic pressure to force a collapse of the magnetic topology.  Therefore, we propose a generalized measure of two-dimensional plasma compressibility:
\begin{equation}
  \Gamma = \frac{b_\perp^2}{2 p + b_{z}^2}
\label{eq:gamma}
\end{equation}
and relate the free parameters of the simulations, $p_0$ and $b_{z0}$, to the magnitude of the in-plane field $b_\perp \approx 1$ ($B_0$), as well as to the initial background plasma $\beta$:
\begin{equation}
  \beta = \frac{2 p_0}{b_\perp^2 + b_{z0}^2}
\label{eq:beta}
\end{equation}

The right panel of Fig.~\ref{fig:sensitivity} provides an alternative way to assess the effect of the initial parameters on the stability of the flux rope and the X-point in terms of dimensionless quantities.  Note that $\beta$ and $\Gamma$ are related to $b_{z}$ such that some combinations of the two are impossible (denoted by gray shading in the figure):
\begin{equation}
  b_z^2 = \frac{(1-\Gamma \beta) b_\perp^2}{\Gamma(1 + \beta)} \ ,
\end{equation}
which implies $\Gamma \beta \le 1$.

Within the accessible parameter space we observe that above a certain level of compressibility (approximately 8, determined empirically), the X-point tends to collapse horizontally and causes the flux rope to descend.  Within the range $4.5 < \Gamma < 8$, the X-point collapses vertically, causing the flux rope to move upwards out of equilibrium (though much more slowly than in a driven eruption).  However, if the plasma is ``stiffened'' beyond a threshold, $\Gamma \lesssim 3$, the fast waves are able to pass through the X-point as their phase speed is no longer close to zero.  Since the waves no longer accumulate at the X-point, they do not cause it to collapse and the equilibrium is preserved.

While it is possible that different perturbations might produce different empirical thresholds of stability, it has not been the goal of the present study to determine particular values but rather to show that the X-point stability can be fundamentally related to the accumulation of fast waves at the X-point, which can be moderated by changing the background compressibility of the plasma.  Similarly, while the magnitude of the dissipative transport coefficients within the visco-resistive MHD simulations can have some impact on the specific stability thresholds via damping of the fast waves emanating from the flux rope, such damping does not qualitatively change the conclusion of this parameter study.

\begin{figure*}[ht]
\centering
\includegraphics[trim=20 10 195 50,clip,height=6.0cm]{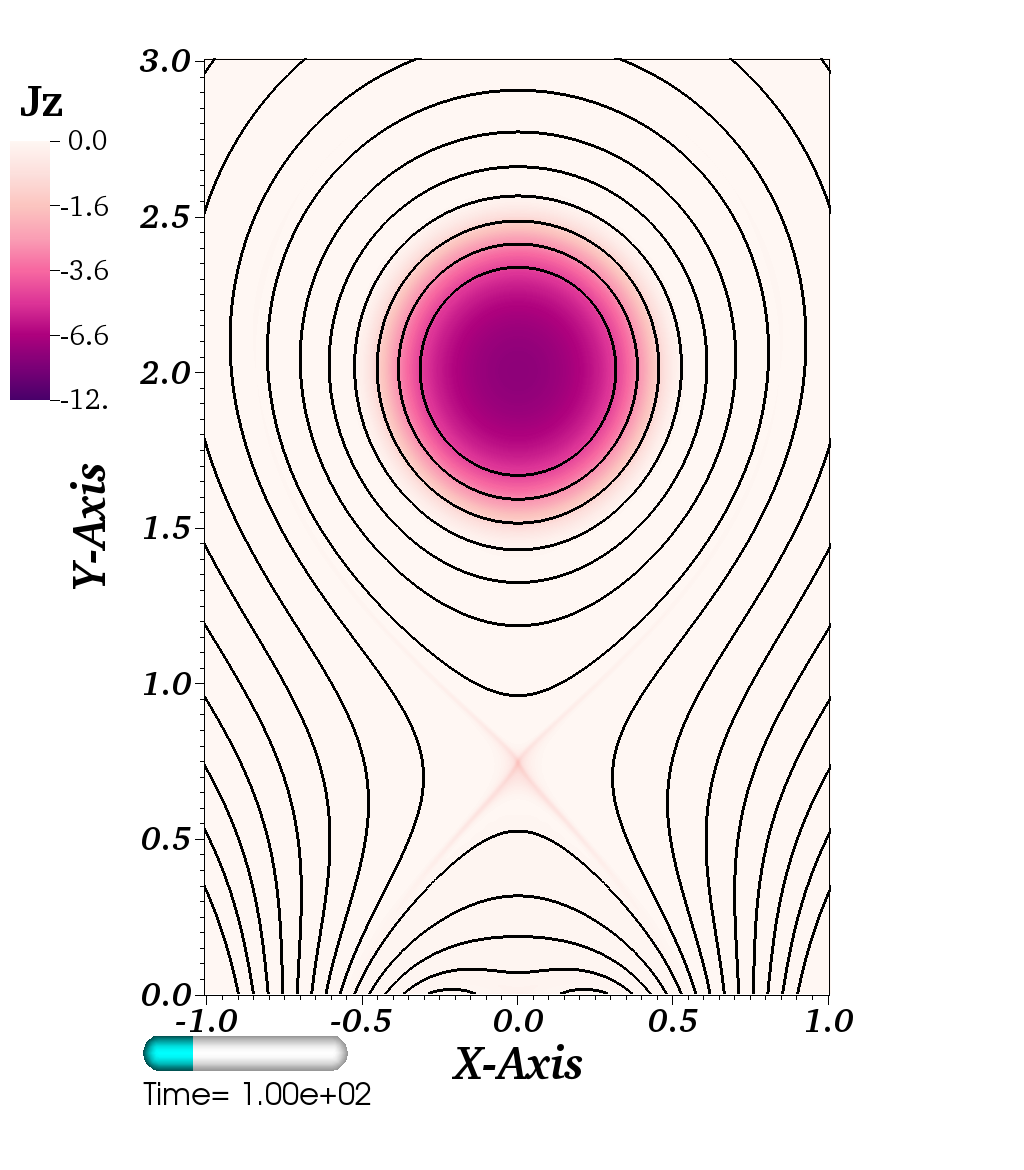} \
\includegraphics[trim=140 10 195 50,clip,height=6.0cm]{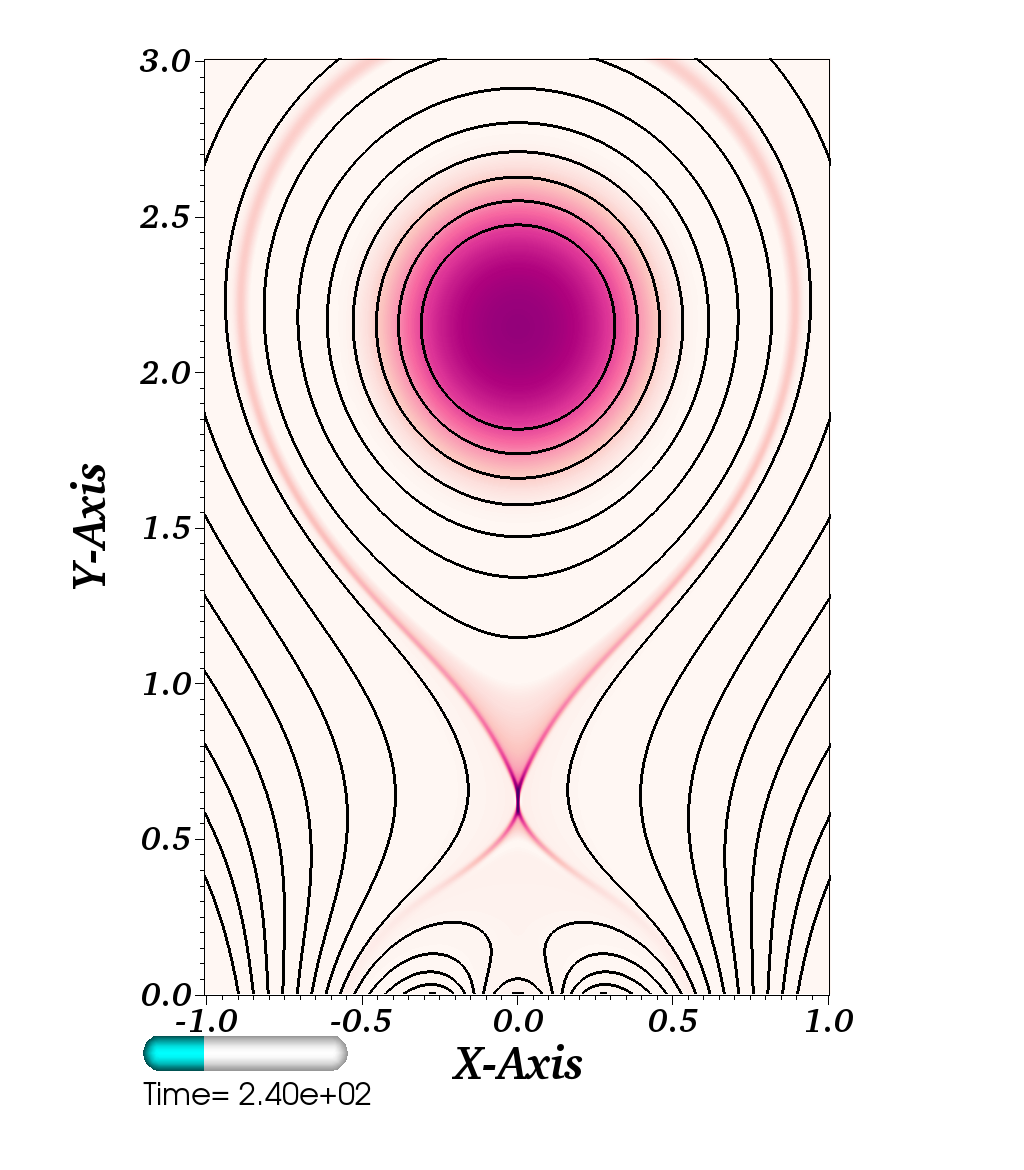} \
\includegraphics[trim=140 10 195 50,clip,height=6.0cm]{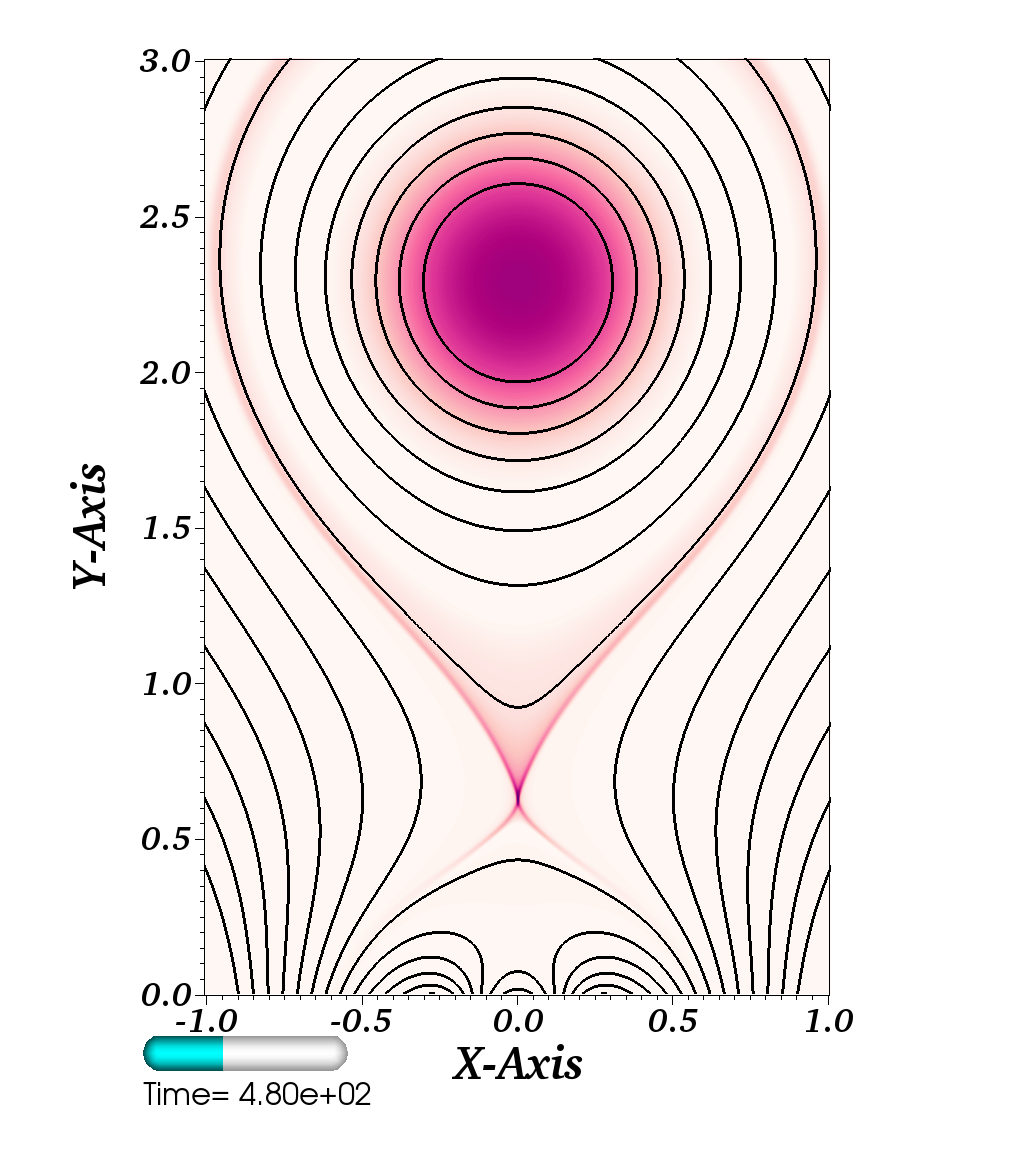} \
\includegraphics[trim=140 10 195 50,clip,height=6.0cm]{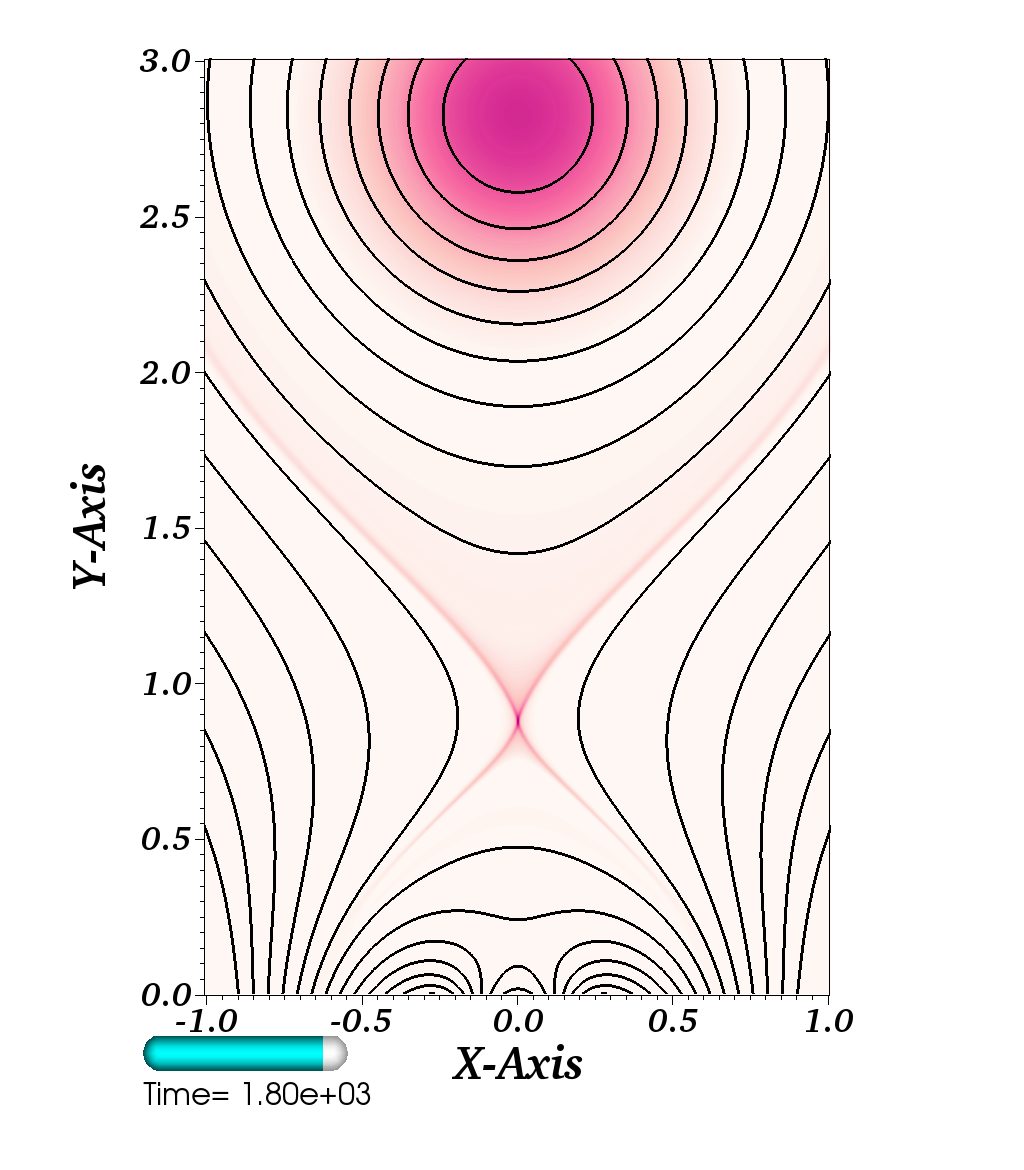} \
\caption{Out-of-plane current density $j_z$ (color, saturated high and low values), with magnetic flux contours (solid black), in a simulation of flux emergence into an unstratified atmosphere.  The four panels show snapshots of the simulation at $100\tau = 12$~min, $240\tau = 29$~min, $480\tau = 58$~min, and $1800\tau = 217.5$~min.}
\label{fig:unstratified_CME_jz}
\end{figure*}
\subsection{CME eruptions driven by flux emergence}
The premise of the CS model is that a stable pre-existing flux rope can be driven to eruption by magnetic flux emergence.  Flux emergence is achieved through photospheric boundary driving (see Eqs.~\ref{eq:driving}): a small amount of flux is effectively emerged through the photospheric boundary by applying a time-dependent electric field.  Emerging flux can cause the flux rope to move in either direction by forcing a destabilization of the X-point, similarly to the fast waves but more predictably.  Within the underlying arcade, when the emergent flux is oppositely oriented to the local flux, it causes a vertical collapse of the X-point, leading to a rising flux rope.  Oriented in the same sense as the local flux rope, it causes a horizontal collapse of the X-point, which forces the flux rope downwards.  For emergence outside the arcade, likewise, it is possible to choose values for the coefficient $c_e$ in Eqs.~\ref{eq:driving} such that the simulation results in a vertical X-point collapse, and when the sign of $c_e$ is reversed, the X-point collapses horizontally.  However, the sign of $c_e$ must be carefully chosen based on topological and geometric considerations, including the sign of the local overlying flux.  In this study, we restrict ourselves to discussing emergence at $x_0=0$ alone, with $c_e=1.1$ as in the original CS model.\footnote{We note that the original CS reference \citet{ChenShibata00} quotes $c_e=11$ and $c=2.5628$, but these should have been quoted as a factor of 10 lower, as per personal communication with P.F.~Chen.}

To evaluate the impact of reconnection micro-physics, stability of the initial condition and atmospheric stratification on the system's response to flux emergence in the CS model, the simulation study described below has been performed by changing one model parameter at a time with otherwise identical numerical and dissipation parameters.  In the reference simulation with an unstratified atmosphere, the background magnetic ``guide" field is set to $b_{z0}=1$, equivalent to $10$~G and of the same order as $b_\perp$, such that the plasma compressibility measure $\Gamma$ is less than unity and the initial configuration is stable for any plasma pressure profile.  To minimize the impact of the size of the computational domain or the dissipative boundary layers on the results of the simulations, the domain boundaries are placed at $L_x=4$ and $L_y=10$.   The computational grid spanning the $(x,y)\in[0,L_x]\times[0,L_y]$ domain has $864$ and $1536$ spatial degrees of freedom in the $x$ and $y$ directions, respectively, distributed non-uniformly in such a way that the vertically elongated X-point reconnection current sheet is well-resolved in the $x$-direction, while both magnetic and thermodynamic structures associated with flux emergence through the chromosphere can be well resolved in the $y$-direction.

The background resistivity throughout the domain is set to $\eta_{bg}=10^{-5}$, the photospheric resistivity is set to $\eta_{ph}=10^{-2}$, there is no anomalous resistivity $\bar{\eta}_{anom}=0$, the background kinematic viscosity coefficient is set to $\mu_{bg}=10^{-4}$, and the heat conduction is set to $\kappa=10^{-5}$.  The duration of the flux emergence is taken to be $t_e=300$, equivalent to $36.25$~minutes.

\begin{figure*}[ht]
\centering
\includegraphics[height=6.2cm]{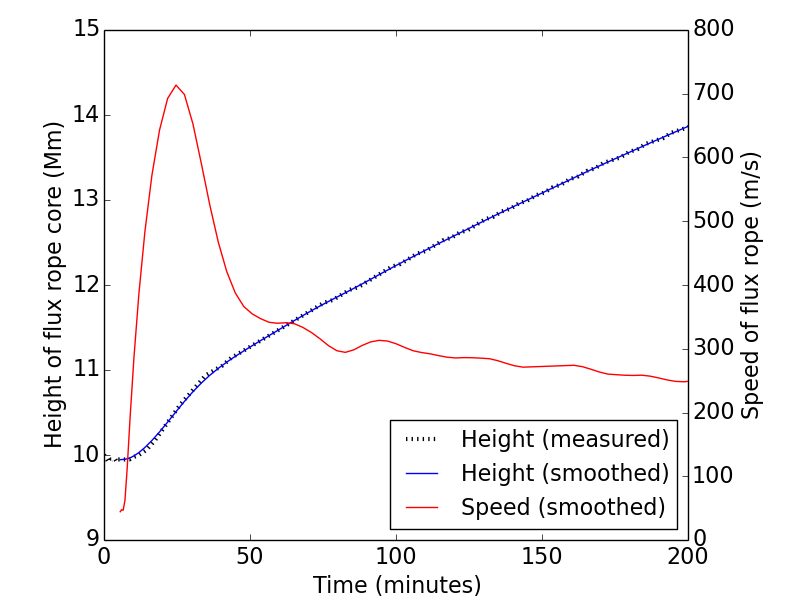} \
\includegraphics[height=6.5cm]{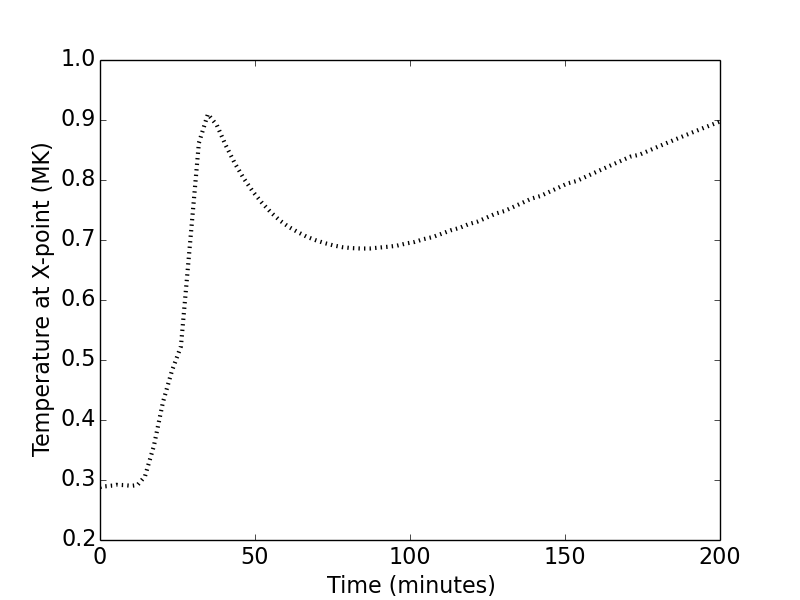} \ 
\caption{{\bf Unstratified atmosphere}.  {\em Left:} Height and speed of flux rope center.  Smoothing is performed using a Hanning window of 12 points.  The speed is computed by finite-differencing the smoothed (blue) curve.  {\em Right:} Temperature at the X-point (center of current sheet) below the flux rope during the same period.}
\label{fig:unstratified_CME}
\end{figure*}
\subsubsection{Flux emergence in an unstratified atmosphere}
To approximate the coronal conditions in the unstratified simulation, the initial pressure is set to $p_0=10^{-2}$, such that the initial $\beta$ is $\sim 1\%$ throughout the domain.  To produce an eruption, the photospheric electric field drive is applied within the arcade below the X-point to generate $B_x$ opposite to the magnetic field of the arcade.   As a result, the magnetic pressure above the photospheric boundary is reduced causing a local downflow towards the photosphere.  This in turn reduces the plasma pressure below the X-point, which forces an in-flow at the sides of the X-point, bringing about its collapse and formation of a reconnection current sheet (\textit{e.g.}, see Fig.~\ref{fig:unstratified_CME_jz}). 

As shown in Fig.~\ref{fig:unstratified_CME_jz}, the X-point collapse  in this simulation is observed to occur at $t\approx 100$, forming a current sheet that reaches its maximum length and strength near $t=200$.  Current density then also increases along the separatrices and the field lines connected to the current sheet.  When the new flux stops emerging ($t > t_e$), the current sheet persists at approximately half to a third of its peak magnitude, slowly diminishing over time for the duration of the simulation.

As reconnection ensues, the flux rope is nudged out of equilibrium (in the $+ \hat{\bvec e}_y$ direction) by the reconnection outflow and continues to move outwards as reconnection proceeds.  The left panel of Fig.~\ref{fig:unstratified_CME} tracks the height of the flux rope center during the eruption by measuring the position of the magnetic O-point (black dots).  The height measurements are smoothed (blue curve) using a Hanning window convolution over 12-point windows, and the speed (red curve) is computed by finite-differencing the smoothed height.  The maximum speed of the flux rope is observed to be only about $0.7$~km/s, quickly slowing down further as the reconnection loses steam.  In the right panel of Fig.~\ref{fig:unstratified_CME}, the temperature at the X-point, or the current sheet center, is plotted in mega-Kelvin showing rapid heating early in the eruption due to Joule heating at the current sheet.

\begin{figure}[ht]
\resizebox{\hsize}{!}
{\includegraphics{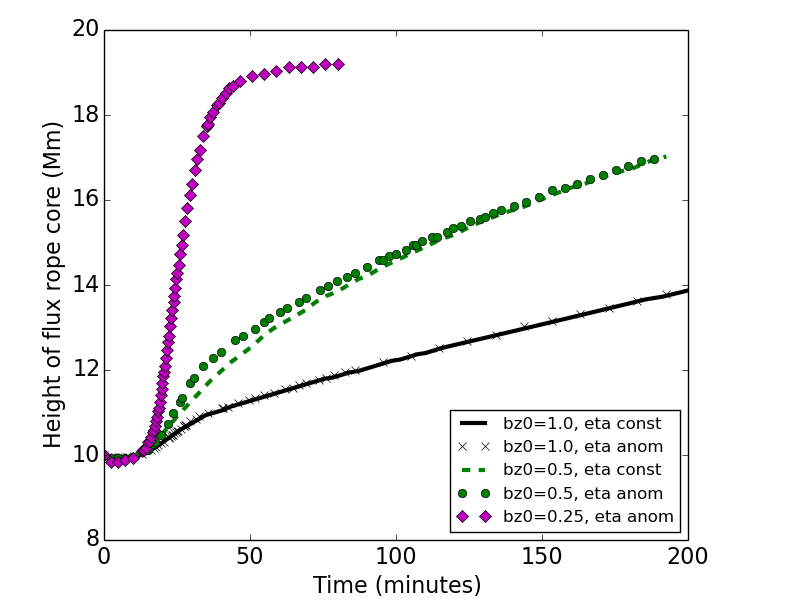}}
\caption{{\bf Unstratified atmosphere}.  Height of flux rope center for a set of five simulations with varying magnitude of initial background magnetic guide field and resistivity models.  Three values of the guide magnetic field $b_{z0}=\{1.0,0.5,0.25\}$ and two resistivity models, one with $\bar{\eta}_{anom}=0$ labeled as ``eta const", and one with $\bar{\eta}_{anom}=10^{-2}$ and $j_c=10$ labeled as ``eta anom", both using the constant background resistivity value $\eta_{bg}=10^{-5}$, are considered.}
\label{fig:unstratified_scan}
\end{figure}

We note that this reference simulation results in a very slowly rising flux rope which is inconsistent with the original \citet{ChenShibata00} simulation where the flux rope rise speed of approximately $70$~km/s was observed.  To study the sensitivity of this result to the magnitude of the background magnetic guide field and the micro-physics of reconnection at the X-point, represented here by the anomalous resistivity model similar to that of \citet{ChenShibata00}, a series of further simulations has been performed.  Figure~\ref{fig:unstratified_scan} shows traces of the height of the flux rope center for a set of five simulations with three different values of the guide magnetic field $b_{z0}=\{1.0,0.5,0.25\}$ and two resistivity models, one with $\bar{\eta}_{anom}=0$ and another with $\bar{\eta}_{anom}=10^{-2}$ and $j_c=10$, both using the constant background resistivity value $\eta_{bg}=10^{-5}$.

The comparison of the five traces clearly demonstrates that the outcome of the simulations is much more sensitive to the magnitude of the background guide field, \textit{i.e.} the global structure and stability of the magnetic configuration, than to the resistivity model.  The two traces with $b_{z0}=1.0$, the initially stable magnetic configuration, are virtually indistinguishable from each other despite very different resistivity models.  The two traces with $b_{z0}=0.5$ initialized from a marginally stable configuration (see Fig.~\ref{fig:sensitivity}) do show small differences during the acceleration phase. Here the simulation with anomalous resistivity allows for slightly faster rise, but both rise much faster than the $b_{z0}=1.0$ cases.  And the initially unstable $b_{z0}=0.25$ case demonstrates yet faster rise of the flux rope that is comparable to the rise speed observed in the \citet{ChenShibata00} simulation.   (Only the anomalous resistivity $b_{z0}=0.25$ simulation trace is shown in Fig.~\ref{fig:unstratified_scan} because the corresponding uniform resistivity simulation produces a very intense X-point current sheet that breaks up due to secondary instabilities \citep{Loureiro07}, leading to formation of further spatial sub-structure which we have chosen not to attempt to resolve.  Detailed investigation of such multi-scale reconnection cases is left for future work.)  We note that the choice of critical current density $j_c=10$ for onset of anomalous resistivity is such that all five simulations achieve $|{\bf j}| > j_c$ at the X-point during the acceleration phase of the flux rope, yet that does not result in significant acceleration of the flux rope for the $b_{z0}=1.0$ and $b_{z0}=0.5$ cases.

It is also of interest that the rapid rise of the flux rope in the $b_{z0}=0.25$ case is followed by stagnation at the height of approximately $19$Mm. Such stagnation is indicative of the system finding a new stable magnetic equilibrium where the upward force on the flux rope is balanced by the magnetic tension distributed throughout the overlying magnetic arcade.
\subsubsection{Flux emergence in a stratified atmosphere}
\label{sssec:stratified}
Introduction of atmospheric stratification, as described in Sec.~\ref{sec:stratified}, leads to a more realistic equilibrium plasma configuration that is much denser at the photosphere than in the unstratified corona-like case.  

\begin{figure*}[ht]
\centering
\includegraphics[height=6.2cm]{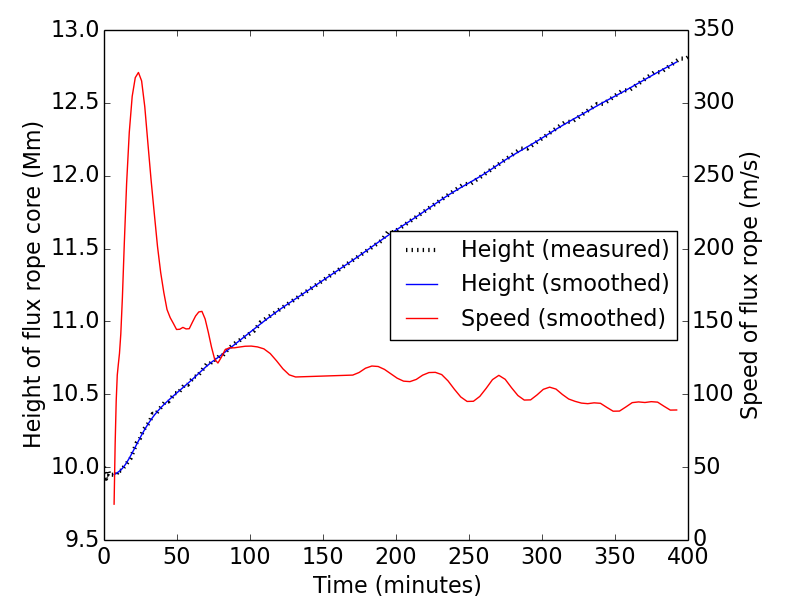} \
\includegraphics[height=6.5cm]{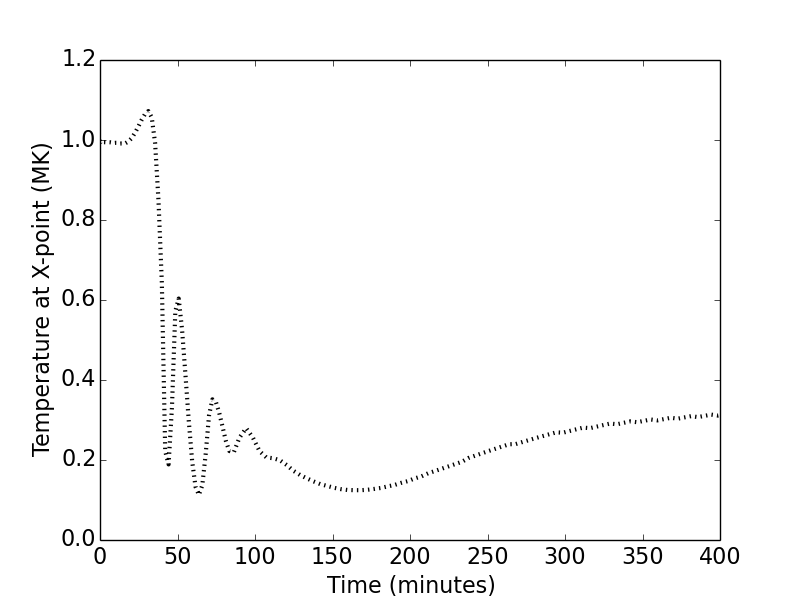} \ 
\caption{{\bf Stratified atmosphere}.  {\em Left:} Height and speed of flux rope center.  Smoothing is performed using a Hanning window of 12 points.  The speed is computed by finite-differencing the smoothed (blue) curve.  {\em Right:} Temperature at the X-point (center of current sheet) below the flux rope during the same period.}
\label{fig:stratified_CME}
\end{figure*}

The impact of the flux emergence at the bottom boundary, with and without the atmospheric stratification, is reflected in the traces of height and speed of the respective flux rope eruptions.  For the stratified atmosphere, the height and speed of the flux rope as functions of time are shown in the left panel of Fig.~\ref{fig:stratified_CME} and can be compared to the equivalent traces for the reference simulation in the left panel of Fig.~\ref{fig:unstratified_CME}. (Note the different ranges of the time axes of the two panels.)  The two time histories are qualitatively similar, both showing rapid acceleration of the flux-rope during flux emergence, with a reduction of the ejection speed by approximately a factor of two once the driving is turned off.  However, both the peak and the post-driving ejection speed of the CME in the stratified atmosphere are less than half of that obtained in the unstratified case.

Another significant difference between the two cases of flux emergence is observed by comparing the time traces of the X-point plasma temperature, shown in the right panels of Fig.~\ref{fig:unstratified_CME} and Fig.~\ref{fig:stratified_CME}. While in the unstratified atmosphere there is a notable temperature increase at the X-point at the time of eruption, in the stratified simulation the temperature decreases instead. Furthermore, as the flux rope begins to rise between $35$ and $110$ minutes into the simulation ($300 \lesssim t \lesssim 900$) the stratified simulation shows an oscillatory X-point temperature as long as the flux rope is within $\approx\:1$~Mm of its original position.  The root cause of the overall X-point cooling can easily be explained as upflows of cold chromospheric plasma being advected into the coronal reconnection region.  Nevertheless, the observed self-induced quasi-periodic oscillatory behavior of the X-point temperature is somewhat unexpected.
\begin{figure*}[ht]
\centering
\includegraphics[trim=10 25 285 50,clip,height=6.5cm]{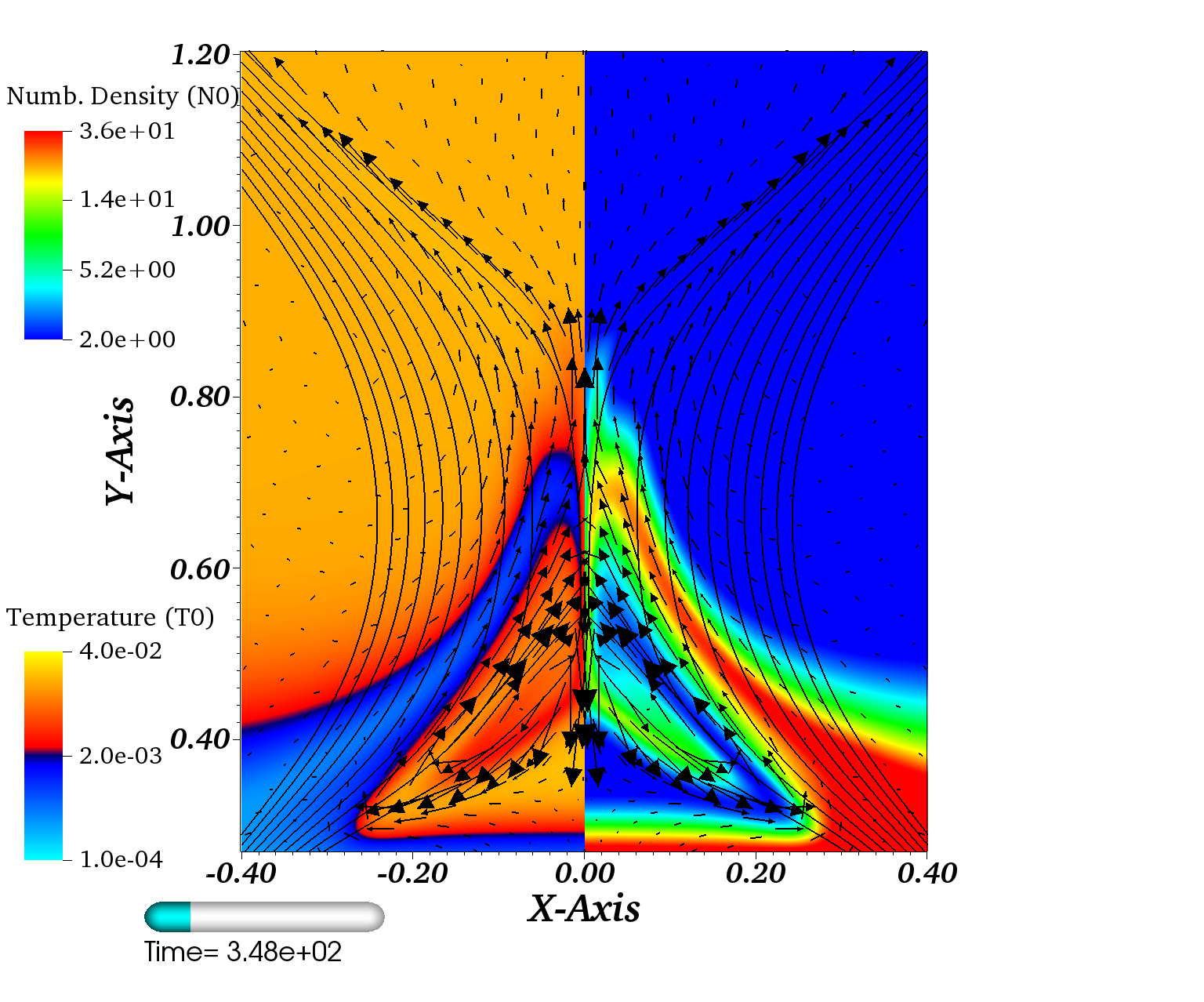} \
\includegraphics[trim=130 25 285 50,clip,height=6.5cm]{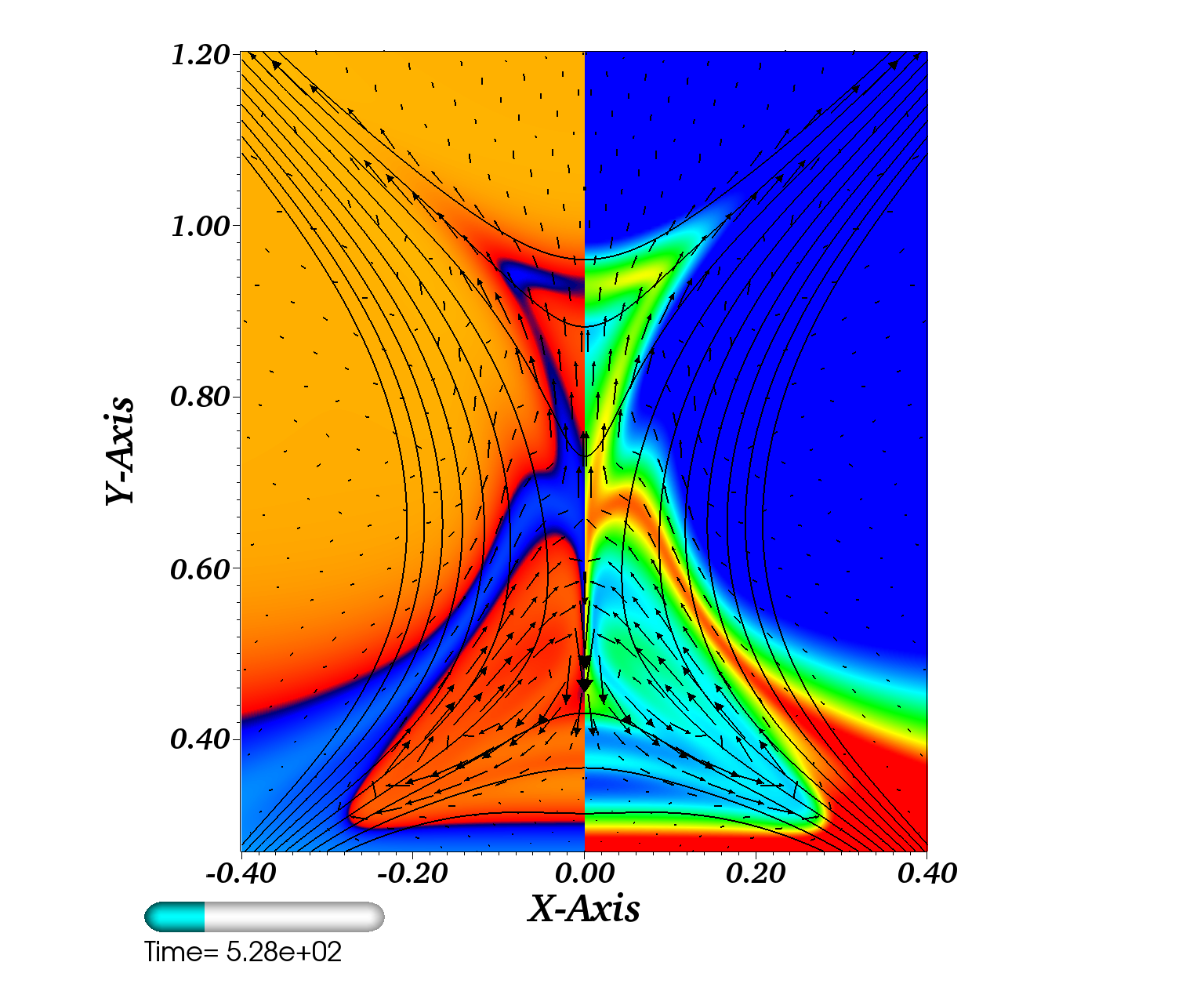} \
\includegraphics[trim=10 25 285 50,clip,height=6.5cm]{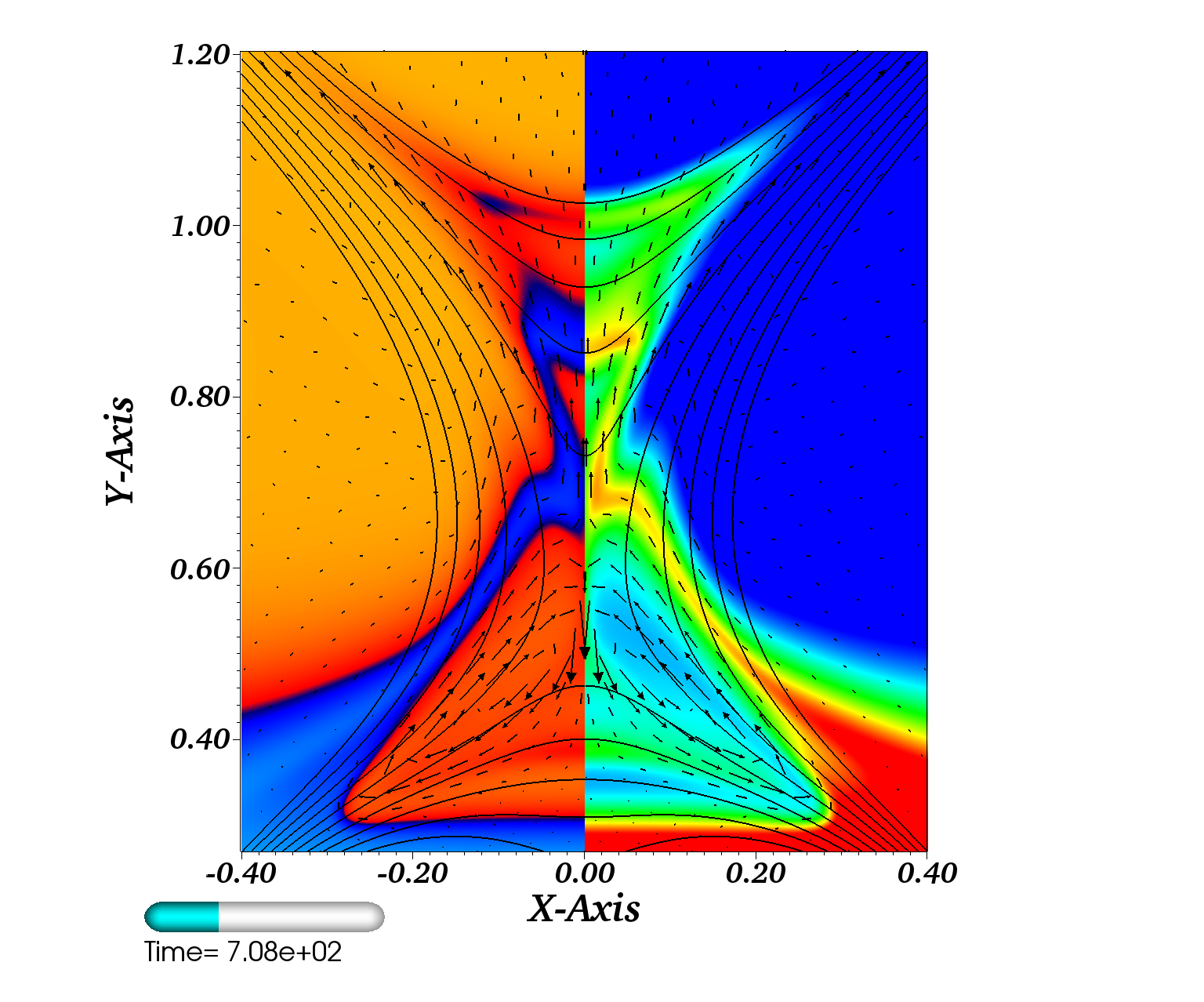} \
\includegraphics[trim=130 25 285 50,clip,height=6.5cm]{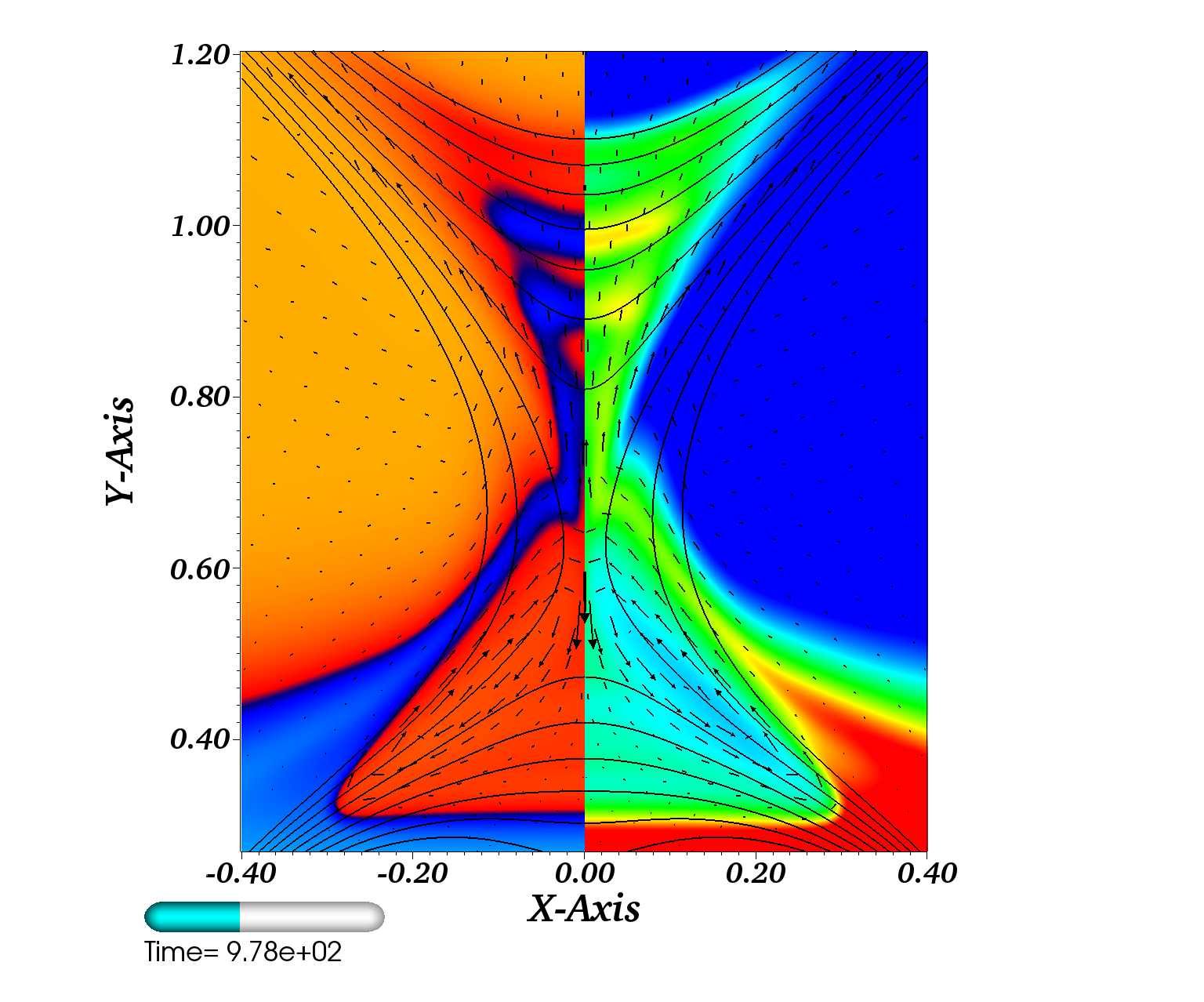} \
\caption{Evolution in time of the reconnection site behind the CME flux rope in a stratified atmosphere.  Each panel shows a snapshot of the temperature structure on the left, the density structure on the right, select contours of the magnetic flux $\psi$ (the same contour values, denoting the same magnetic field lines, have been chosen for each panel), and arrows denoting the in-plane plasma flow. The snapshots are made $348\tau = 42$~min, $528\tau = 64$~min, $708\tau = 86$~min, and $978\tau = 118$~min into the simulation.  Note that for illustration purposes both plasma temperature and number density are plotted using logarithmic color scales with saturated high and low values.  Arrows showing the plasma flow have been scaled by a factor of 25 with respect to the linear dimensions of the domain so that an arrow of unit length corresponds to flow of $1.7\times 10^4$~km/s.}
\label{fig:stratified_Xpoint}
\end{figure*}

Fig.~\ref{fig:stratified_Xpoint} shows the evolution in time of plasma temperature, density, and flows around the X-point during the period of quasi-periodic temperature oscillations.  Continuous upflows of dense cool plasma convected along the magnetic field lines and into the reconnection region around the X-point are apparent throughout the evolution.  The lower-right panel of this figure makes clear that this continuous chromospheric upflow results in quasi-periodic striations of cool dense material alternating with hotter, lower-density plasma on the recently reconnected field lines rising towards (and with) the flux rope located above.  These striations are the signatures of the same oscillatory behavior observed on the X-point temperature trace in Fig.~\ref{fig:stratified_CME}.
While the origins and parametric robustness of the observed quasi-periodic phenomenon require further in-depth study that is outside of the scope of this article, a heuristic explanation of the basic physical mechanism is straightforward.  It results from the competition between the upward directed tension force in newly reconnected magnetic field lines and the downward directed gravity acting on the dense, cold plasma deposited onto these same field-lines by the chromospheric upflows.  As in the formation of water droplets, whenever sufficient amount of plasma accumulates in a small enough volume in the V-shaped dip of a set of recently reconnected field lines, the gravitational pull on that plasma overcomes the field's tension force and a droplet of plasma forms and falls vertically through the reconnection site itself. As a result, those flux-rope destined field lines that produce the droplets end up with lower density hotter plasma, while the field lines that pass through in between the droplets contain colder and heavier plasma.  The temperature at the X-point, where the reconnection is regularly disrupted by the droplets, is similarly modulated when the plasma that has been heated by the reconnection process is periodically replaced by the cold plasma of the droplets. 

Below the reconnection site, the pattern of chromospheric upflows along the magnetic separatrices and vertical downflows through the X-point creates a circulation of plasma between reconnection's outflow and inflow.  How, and whether or not, this circulation pattern contributes to the formation of the quasi-periodic temperature and density structure described above is left as a topic for future study.
\section{Discussion \& Conclusions}
\label{sec:discussion}

Coronal mass ejections are eruptive solar events of enormous proportions that shed plasma and magnetic flux into interplanetary space.  The Chen \& Shibata model is a good starting point for understanding how such an eruption can originate from the destabilization of a global magnetic configuration by local flux emergence.  It helps us to see a connection between flux emergence, a phenomenon at the solar surface, and flux rope ejection, a phenomenon in the corona.  Many observational studies have shown spatio-temporal correlations between flux emergence and eruptive events, but few theoretical models to date have identified a precise single mechanism or sequence of processes whereby producing magnetic flux at the photosphere dynamically triggers an eruption.  The CS model may assume an oversimplified solar atmosphere and a somewhat manufactured magnetic topology, but it does proffer a complete story.  To determine the effects of a more realistic solar atmosphere, we have undertaken an effort to repeat the study using a different numerical suite and allowing for a stratified atmosphere with the density variation of over four orders of magnitude, as well as a sharp temperature transition between the chromosphere and the corona.  

We have found that even in the absence of stratification the initial equilibrium can be unstable to small perturbations.  The initial adjustment of the magnetic equilibrium to slight force imbalances can generate fast waves that may not be able to propagate through the X-point below the flux rope.  In these cases, the fast waves accumulate in such a way as to collapse the X-point and initiate reconnection.  Thus, the equilibrium can be destabilized before any photospheric driving is applied.  However, we also found that the stability of the CS equilibrium can be controlled by varying the compressibility of the plasma, which in a two-dimensional system is determined by the combination of thermal pressure and the magnitude of the out-of-plane component of the magnetic field.  To quantify this effect, we defined a generalized measure of compressibility $\Gamma$ and have empirically determined the equilibrium's stability boundaries in terms of $\Gamma$.  

When emulating flux emergence by applying an electric field at the photospheric boundary, in the unstratified atmosphere, the results of our simulations are qualitatively similar to those of the original study.  However, there are also important differences and new findings.  As opposed to the original study, when initialized in a stable configuration, our simulations show little evidence of significant flux rope acceleration or Joule heating associated with the reconnection current sheet.  Notably, this result appears to be insensitive to the micro-physics of the reconnection region. By varying the magnitude of the background out-of-plane magnetic field component and thus changing the stability of the global magnetic configuration, we also show that flux rope rise speeds comparable to the original result are possible but require an unstable magnetic configuration as the initial condition.

We further show that the micro-physics of reconnection is more likely to slow down than to accelerate the flux rope by comparing simulations with and without anomalous resistivity.  It is well known that current-dependent anomalous resistivity allows for ``fast" magnetic reconnection with only weak dependence on the magnitude of resistivity itself \citep{Malyshkin05}.  Yet, for both initially stable and quasi-stable magnetic configurations, allowing for anomalous resistivity did not result in a substantial increase of the flux rope rise speed.  That is, merely allowing for faster reconnection did not lead to faster reconnection and faster flux rope ejection.  On the other hand, in magnetic configurations where fast flux-rope ejection is possible, the simulations with low guide field indicate that the inability of the magnetic reconnection process to occur sufficiently fast could limit the rise speed of the flux rope.

In the flux emergence simulations with stable magnetic configuration and realistic atmospheric stratification, the weakness of the X-point heating and the slowness of the ejected flux rope are reproduced, and amplified.  In these simulations, changes in the magnetic field structure due to flux emergence generate persistent chromospheric upflows of cold, dense material that is convected into and dramatically cools the reconnection current sheet.

In addition to the steady state upflows and cooling, the stratified simulations also produce another type of behavior: self-induced quasi-periodic oscillations in the X-point temperature, density, and other fluid quantities.  The quasi-periodic oscillations observed in the stratified simulation are of transient nature, appearing after the flux emergence drive has been completed and lasting for just over an hour while the flux rope is within $\approx 1$~Mm of its initial location.  The robustness of this phenomenon will be a subject of future research, but our initial investigation indicates that a critical balance between the upward tension force of the reconnected magnetic field and the downward gravitational pull on the dense chromospheric plasma convected into the reconnection region has to be achieved in order for the quasi-periodic oscillations to appear in a simulation.  While that may seem to be a prohibitive constraint, we speculate that in the three-dimensional parameter space spanned by (1) the height of the X-point, (2) the strength of the magnetic fields and (3) the horizontal location of the emerging flux relative to the separatrices of the pre-existing magnetic configuration, all quantities that can vary greatly throughout the lower solar atmosphere, there is likely embedded a two-dimensional parameter space where such balance can, indeed, be achieved.

We note that there is also extensive observational evidence for what has been called quasi-periodic pulsations (QPP) in solar and stellar flares \citep[e.g., see][and references therein]{Nakariakov09, MitraKraev05} with the QPP periodicity time scale varying from fractions of a second to several minutes, comparable to the period of the oscillations produced in our simulation.  In fact, \citet{Nakariakov09} have previously resorted to the water drop formation analogy in describing what they refer to as a class of ``load/unload'' models of long multi-minute period QPPs.  The plasma droplet mechanism described in Sec.~\ref{sssec:stratified} above is a much more direct, and novel, analogy to the same physical process with the potential to provide a new alternative explanation for the long-duration QPPs.

Finally, we point out that the limitations of the two-dimensional MHD model used here for modeling a region of potential flaring activity embedded into a stratified solar atmosphere are many.  It is well known that laminar resistive reconnection cannot account for the observed rates of magnetic energy release, particle acceleration, or radiation from solar flares, while three-dimensional effects can substantially alter both the flux-rope stability properties and the micro-physics of reconnection. 
Nevertheless, we believe that the careful and systematic study described in this article is a prerequisite for performing more complete, and also substantially more challenging and complicated, studies of CME initiation by flux emergence in the future. 

\begin{acknowledgements}
E.L.~thanks Neil Sheeley for his valuable insights into solving the hyperbolic function integrals.  This work was supported by the NASA SR\&T and LWS programs, as well as ONR 6.1 program.  Simulations were performed under grants of computer time from the US DOD HPC program and the National Energy Research Scientific Computing Center, which is supported by the US DOE Office of Science. 
\end{acknowledgements}

\section*{Appendix A}  
We derive the uniform pressure magnetic flux rope equilibrium with axial field from a familiar form of the Grad--Shafranov equation:
\begin{equation}
  \frac{d}{d\psi}\left(\frac{b_z^2}{2}\right) = -\nabla^2 \psi = j_z
\end{equation}
In particular, assuming $\psi(r)=\psi_l(r)$ given by Eq.~\eqref{eq:psi_l1}:
\begin{align}
  \frac{b_z^2}{2} &= \int j_z \;d\psi = \int j_z \frac{d\psi}{dr} dr \nonumber \\
                  &= \int \left[ \frac{4r^2}{r_0^3} - \frac{4}{r_0} \right] \left[\frac{r}{r_0} - \frac{r(r^2-r_0^2)}{r_0^3}\right] dr  \\
                  &= -\frac{2}{3}\left(\frac{r}{r_0}\right)^6 + 3\left(\frac{r}{r_0}\right)^4 - 4 \left(\frac{r}{r_0}\right)^2 + \text{constant,}\, 
                  \text{for}\ r \le r_0 .\nonumber
\end{align}
Requiring that $b_z = b_{z0}$ for $r \ge r_0$, we can determine the constant of integration such that $b_z$ is continuous:

\begin{equation}
  b_z(r) = \left\{ 
    \begin{array}{l @{\hspace{6mm}} c}
       \sqrt{b_{z0}^2 + \dfrac{10}{3} - 8 \left(\dfrac{r}{r_0}\right)^2 + 6  \left(\dfrac{r}{r_0}\right)^4 -\dfrac{4}{3} \left(\dfrac{r}{r_0}\right)^6} \, & r \le r_0 \\[3mm]
       b_{z0} \ . & r > r_0
    \end{array}
  \right.
\end{equation}

Suppose, however, we wish to find $b_z$ as a function of $\psi$, rather than of $r$.  We then solve for the inverse function $\zeta \equiv \psi_l^{-1}(r)$ by replacing $r$ with $\zeta$ in Eq.~\eqref{eq:psi_l1} and rearranging terms:
\begin{equation}
\zeta^4 - 4r_0^2\zeta^2 + r_0^4 + 4 r_0^3 \psi_l = 0 \ .
\end{equation}
Solving this quadratic equation for $\zeta^2$, we find:
\begin{equation}
  \zeta^2 = 2 r_0^2 \pm \sqrt{3r_0^4 - 4 r_0^3 \psi_l } \ .
\end{equation}
We recover the form of Eq.~\eqref{eq:rstar_approx} by rejecting the positive root (to permit small values of $\zeta$), replacing $\psi_l$ by the full functional form of $\psi = \psi_l + \psi_i + \psi_b$ to approximate a force-free initial condition with well-aligned contours of constant $\psi$ and $b_z$, and allowing for gauge freedom.

\section*{Appendix B}  
Here we present a derivation of the pressure profile used in simulations of a stratified solar atmosphere, given a particular temperature profile \eqref{eq:temperature}.  To be physically relevant, we use here dimensional quantities, rather the normalized code variables.

We begin with the first-order differential equation governing hydrostatic equilibrium:
\begin{equation}
  \frac{dp}{dy} + m_p n g_S = 0 \ ,
\label{eq:hydrostatic2}
\end{equation}
which we divide by $p=2 n k_B T$:
\begin{equation}
  \frac{d \ln p}{dy} + \frac{m_p g_S}{2 k_B T} = 0.
\end{equation}
Then
\begin{equation}
  \ln \frac{p}{p_0} = -\frac{m_p g_S}{2 k_B} \int \frac{dy}{T}.
\label{eq:p_integration}
\end{equation}
We use the profile for temperature $T$ given by \eqref{eq:temperature}, but with the following variable substitution:
\begin{equation}
 u \equiv \frac{y-y_\text{\tiny TR}}{\Delta y} \ ,
\end{equation}
leading to
\begin{equation}
  \begin{split}
    T(u) &= T_p + \frac{T_c-T_p}{2} \left(1 + \tanh u\right) \\
       &= T_p + \frac{T_c-T_p}{2} \left(1 + \frac{e^u-e^{-u}}{e^u+e^{-u}} \right) \\
       &= \frac{T_p\, e^{-u} + T_c\, e^u}{e^u + e^{-u}}.
  \end{split}
\label{eq:tempr}
\end{equation}
With algebraic manipulations, we can rewrite \eqref{eq:tempr} as:
\begin{equation}
  \frac{1}{T} = \frac{1}{T_c} + \frac{e^{-2u}(1-T_p/T_c)}{T_p\,e^{-2u} + T_c}.
\end{equation}
Then the integral in \eqref{eq:p_integration} can be evaluated:
\begin{equation}
  \begin{split}
    \int \frac{dy}{T} &= \Delta y \int \frac{du}{T} = \Delta y \left[ \int \frac{du}{T_c} + \left( 1-\frac{T_p}{T_c} \right) \int \frac{e^{-2u}\,du}{T_p\,e^{-2u}+T_c} \right] \\
    &= \Delta y \left[ \frac{u}{T_c} - \frac{1}{2}\left(\frac{1}{T_p}-\frac{1}{T_c}\right) \ln\left(T_p e^{-2u} + T_c\right) \right].
  \end{split}
\label{eq:T_integration}
\end{equation}

Finally, substituting \eqref{eq:T_integration} into \eqref{eq:p_integration} yields an expression for $p$, in terms of $u$:
\begin{equation}
  p(u) = p_0 \exp\left\{ \frac{m_p g_S\Delta y}{2 k_B T_c} \left[\frac{T_c-T_p}{2 T_p} \,\ln \left(T_p e^{-2u} + T_c\right) - u \right] \right\}.
\end{equation}

\small
\bibliography{bibliography}

\end{document}